\documentclass[12pt,preprint]{emulateapj}

\usepackage{natbib}

\newcommand{\degree}{\ensuremath{^\circ}}
\newcommand{\curlym}{\ensuremath{\mathcal{M}}}
\newcommand{\fkoz}{\ensuremath{f_{\rm koz}}}
\newcommand{\fkoztrue}{\ensuremath{f_{\rm koz}^{\rm true}}}
\newcommand{\fkozobs}{\ensuremath{f_{\rm koz}^{\rm obs}}}
\newcommand{\imax}{\ensuremath{I_{\rm max}}}

\begin{document}

\title{Plutino Detection Biases, Including the Kozai Resonance}

\author{S.~M. Lawler\altaffilmark{1},
B. Gladman\altaffilmark{1}
}

\altaffiltext{1}{Department of Physics and Astronomy, 6224
  Agricultural Road, University of British Columbia, Vancouver, BC V6T 1Z1}

\begin{abstract}

Because of their relative proximity within the transneptunian region, 
the plutinos (objects in the 3:2 mean-motion resonance with Neptune) are 
numerous in flux-limited catalogs, and well-studied theoretically.
We perform detailed modelling of the on-sky detection biases for plutinos, 
with special attention to those that are simultaneously in the Kozai resonance.  
In addition to the normal 3:2 resonant argument libration,
Kozai plutinos also show periodic oscillations in eccentricity and inclination, 
coupled to the argument of perihelion ($\omega$) oscillation.  
Due to the mean-motion resonance, plutinos avoid coming to pericenter near 
Neptune's current position in the ecliptic plane.  
Because Kozai plutinos are restricted to certain values of $\omega$, 
perihelion always occurs out of the ecliptic plane, biasing 
ecliptic surveys against finding these objects.  
The observed Kozai plutino fraction $\fkozobs$ has been measured 
by several surveys, finding 
values between 8\% and 25\%, while the true Kozai plutino fraction 
$\fkoztrue$ has been predicted 
to be between 10\% and 30\% by different giant planet migration simulations.
We show that $\fkozobs$ varies widely depending on the 
ecliptic latitude and longitude of the survey, so debiasing to find the 
true ratio is complex.  
Even a survey that covers most or all of the sky will detect an apparent Kozai 
fraction that is different from $\fkoztrue$.  
We present a map of the on-sky plutino Kozai fraction that would be detected by 
all-sky flux-limited surveys.  
This will be especially important for the 
Panoramic Survey Telescope \& Rapid Response System (Pan-STARRS)
and Large Synoptic Survey Telescope (LSST) projects, which 
may detect large numbers of plutinos as they sweep the sky.
$\fkoztrue$ and the distribution of the orbital elements of Kozai plutinos may be
a diagnostic of giant planet migration; future migration simulations
should provide details on their resonant Kozai populations.

\end{abstract}

\section{Introduction}

Plutinos are trans-Neptunian objects (TNOs) that are in the 3:2 mean-motion resonance 
with Neptune, meaning that in the time it takes Neptune to complete three orbits of the Sun,
plutinos complete two orbits.
Plutinos are named after Pluto, which was the first known resonant TNO,
discovered to be so by examination of a numerical orbital integration \citep{CohenHubbard1964}.
This same technique is currently used to confirm the resonant nature of
plutinos \citep[and other resonant TNOs;][]{Gladmanetal2008, LykawkaMukai2007}.
Several of the first TNOs that were discovered in the early 1990s were also found to be
plutinos \citep{Daviesetal2008}.
Plutinos are among the easier TNOs to observe;
their semimajor axis of approximately 39.5~AU places them near
the inner edge of the Kuiper Belt.
Discovery is also helped by the high eccentricities many plutinos possess,
causing close-in pericenters and lower apparent magnitudes, and thus brighter objects
(this is an extremely strong effect since the TNOs are observed with reflected light,
so their flux is proportional to distance$^{-4}$).
Plutinos (and other resonant TNOs) can have lower perihelia 
than non-resonant TNOs 
because the mean-motion resonance protects them
from close encounters with Neptune \citep[e.g.][]{Malhotra1996}.

As of November 2012 there were 244 plutinos listed in the Minor Planet Center (MPC) database.
However, only about 120 of these have high-quality multi-opposition orbits, allowing 
numerical orbital integrations \citep{LykawkaMukai2007,Gladmanetal2008}
to prove that the objects are truly resonant
(showing libration of the resonant angle; see Section~\ref{sec:resdynamics}), and not just
located near the resonance phase space.

This paper will highlight the special selection effects caused by the 
Kozai resonance within the 3:2 resonance.
``Kozai plutinos'' are TNOs that are simultaneously in the 3:2 mean-motion resonance with 
Neptune and in the Kozai resonance.  
Pluto was the first known Kozai resonator \citep{WilliamsBenson1971}.

What is now known as the Kozai resonance was first 
described for high-inclination asteroids by \citet{Kozai1962}.   
This effect occurs when a small body in orbit around a large central mass is perturbed by 
a third mass at high relative orbital inclination $i$.
In the case of the trans-Neptunian region, the small bodies are the TNOs themselves, and 
the high relative inclination perturber is Neptune, and to a smaller degree the other planets.  
Any gravitational perturbations on the orbit of a small body cause $\omega$ to change.
Most of the time, this causes $\omega$ to precess, but under the special conditions
that the perturber is at high relative inclination, $\omega$ will oscillate instead: 
this is the signature of the Kozai resonance.
When `near' the resonance in phase space, the $\omega$ evolution becomes highly non-uniform
(sometimes called the Kozai effect rather than resonance).
In either case, one observes coupled $e$ and $i$ oscillations correlated with the 
value of $\omega$.
For non-resonant TNOs, \citet{ThomasMorbidelli1996} show that the Kozai resonance
only occurs at extremely large eccentricity.
However, inside mean-motion resonances, the Kozai 
resonance can appear at moderate $i$ \citep[e.g.][]{Gallardoetal2012}.
This is how Pluto can show Kozai oscillations despite being at the relatively low inclination
of 17$\degree$.

\citet{LykawkaMukai2007} presented the largest collection of plutinos that has been analyzed 
for Kozai resonance, after combing the contents of the Minor Planet Center (MPC) database at the time. 
They found that 22 out of 100 plutinos were solidly in the Kozai resonance, with $\omega$ oscillating 
around 90$\degree$ or 270$\degree$, and in one case around 0$\degree$.
Unfortunately, the MPC database's collection of detections from many different non-uniform
surveys does not allow for easy debiasing.
Because of this, it is very difficult to measure the true fraction of Kozai versus non-Kozai plutinos
from this dataset.

This manuscript was motivated by the results of the 
Canada-France Ecliptic Plane Survey  
\citep[CFEPS;][]{Jonesetal2006, KavelaarsL3, PetitL7, GladmanReson}, which detected and then re-acquired nearly 
200 TNOs, including 24 plutinos, 
to produce high-quality orbits and orbital classifications.
This survey was well calibrated: tracking efficiencies and magnitude depths are
precisely known for each pointing.  
Because of this, the survey could be debiased to produce absolute populations.
Two of the 24 detected Plutinos in the survey were found to be in the Kozai resonance upon
examination of a 10~Myr orbital integration.  
Because 8\% of the CFEPS-discovered plutinos were Kozai plutinos, 
in order to properly debias CFEPS's result to produce the absolute population
and orbital element distribution of the plutinos, 
this Kozai component had to be included and properly modelled \citep{GladmanReson}.

Though this paper only discusses plutinos in the Kozai resonance,
\citet{LykawkaMukai2007} also catalogued Kozai resonators in the 5:3, 7:4, and 2:1
mean-motion resonances.
Other resonances can also exhibit Kozai oscillations in some portion of the 
resonant phase space.

This paper provides a basic understanding of the Kozai resonance 
within the 3:2, how the Kozai dynamics affect plutino observations,
and introduces possible uses for this resonance in distinguishing between different giant
planet migration models.
Section~\ref{sec:resdynamics} discusses in-depth the dynamical requirements for 
a TNO to be in a mean-motion resonance. 
Section~\ref{sec:nonkoz} presents toy models of the 3:2 resonance
(Sections~\ref{sec:pluttoymodel0} and \ref{sec:pluttoymodel95}),
in order to demonstrate the effect of libration,
and finally gives a realistic plutino distribution (Section~\ref{sec:plutrealdist}).
Section~\ref{sec:kozplutinos} discusses both the dynamics of the Kozai
plutinos and the on-sky detection biases
that result from the dynamical constraints placed on the Kozai plutinos.
Section~\ref{sec:pop} gives a summary of how we simulate the plutino population.
This simulation is used in Section~\ref{sec:biases} to show biases that observers
will encounter on different parts of the sky in detecting Kozai and non-Kozai plutinos.
Section~\ref{sec:fkoz} gives a discussion of previous observations of 
Kozai plutinos and of theoretical predictions in the literature.
And finally, Section~\ref{sec:conclusion} discusses future observations
that may help constrain the Kozai fraction and the distribution
of the orbital elements of Kozai plutinos.

\section{Resonant Dynamics} \label{sec:resdynamics}

\begin{figure}
\centering
\includegraphics[scale=0.2]{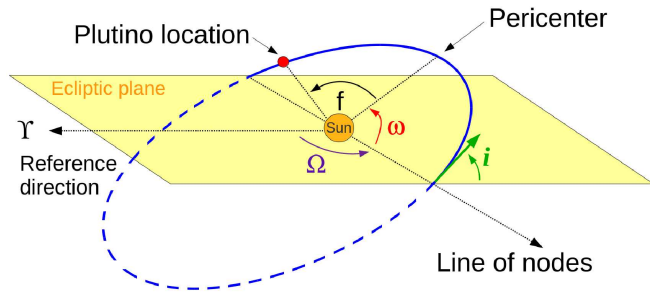}
\caption{
$\Omega$ is the longitude of the ascending node, $\omega$ is the argument of pericenter,
and $f$ is the true anomaly.
The mean anomaly $\curlym$ is the time since the last pericenter times the mean motion: 2$\pi$/P.
The mean longitude $\lambda=\Omega+\omega+\curlym$,
and the longitude of pericenter $\varpi=\Omega+\omega$.
}
\label{fig:orbelems}
\end{figure}

Here we will discusses Plutino dynamics, and how resonant objects are
identified in orbital integrations.
Much of this discussion can be generalized to other mean-motion resonances.
Figure~\ref{fig:orbelems} defines the usual heliocentric ecliptic orbital elements.

\begin{figure*}
\centering
\includegraphics[scale=0.26]{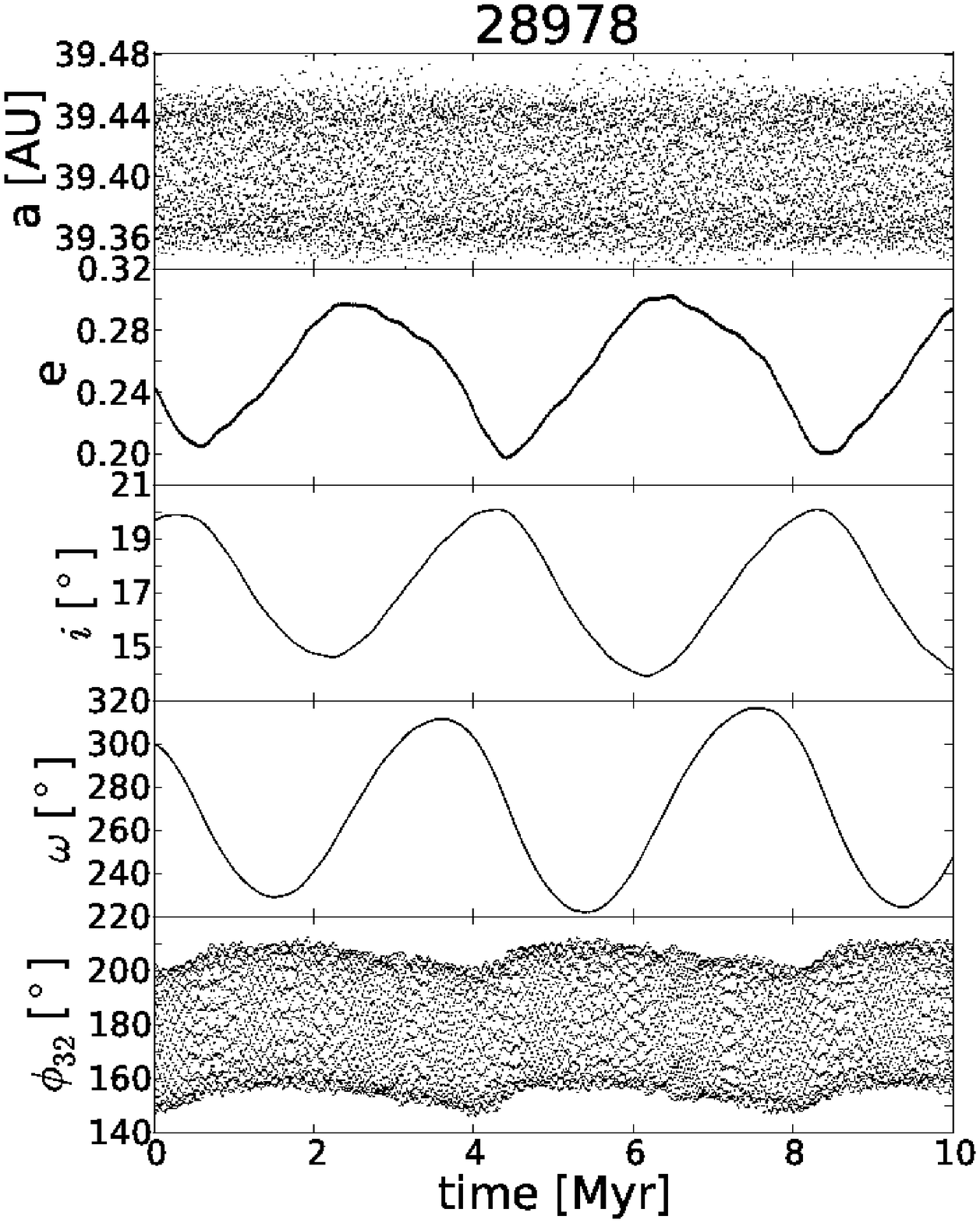} \includegraphics[scale=0.26]{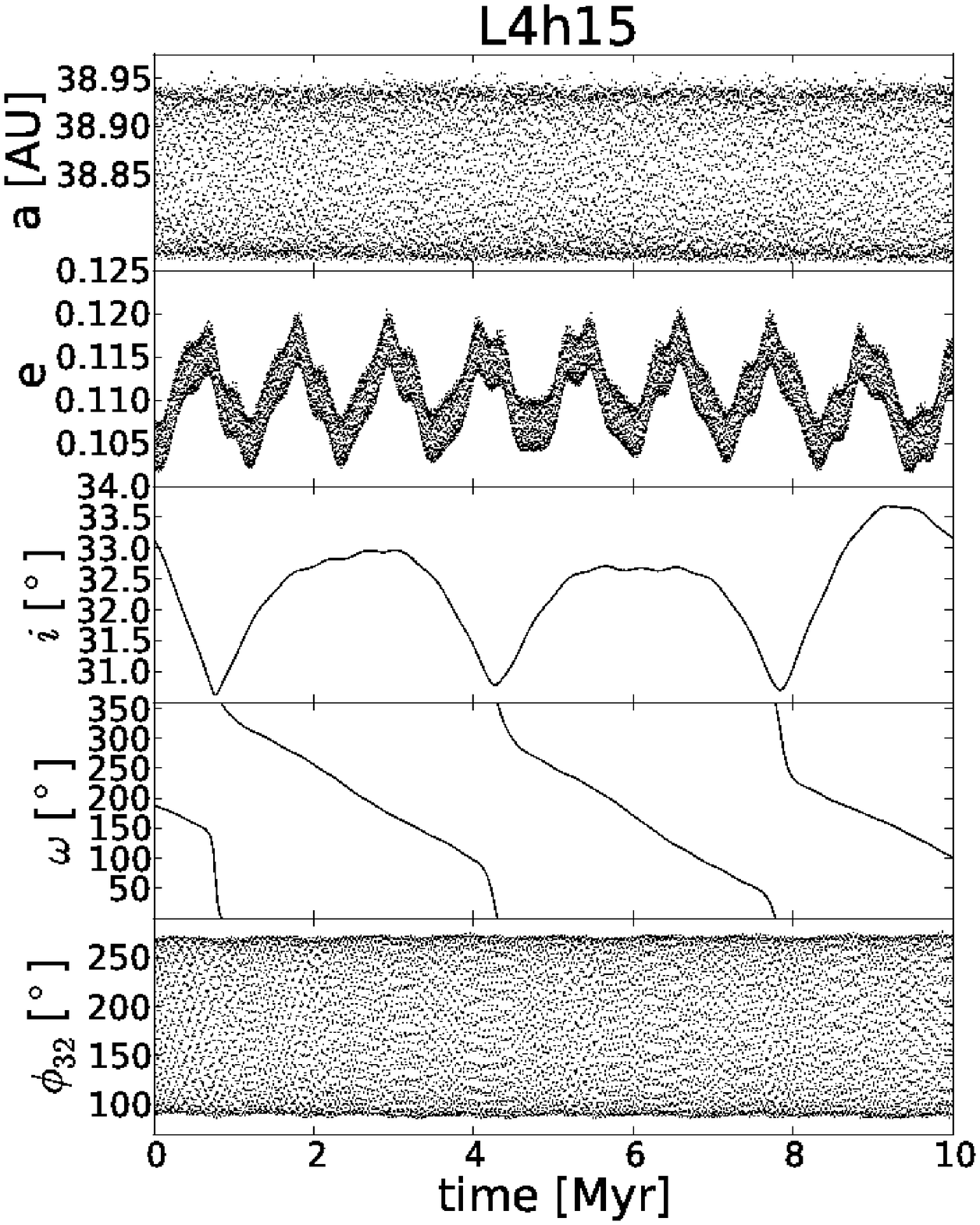} \includegraphics[scale=0.26]{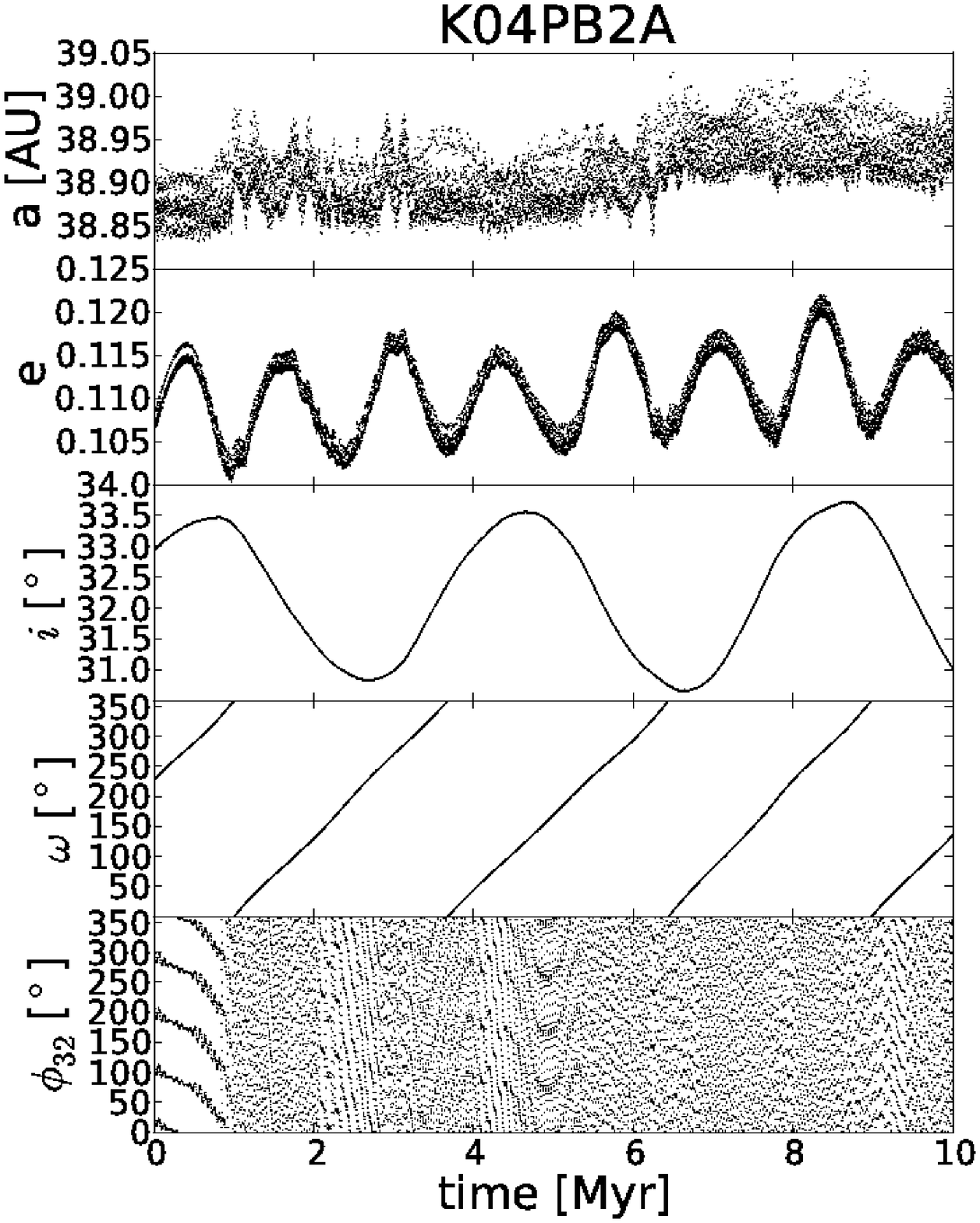}
\caption{Barycentric orbital elements over the course of 10~Myr orbital integrations of the Kozai plutino 28978 Ixion (left),
the CFEPS-discovered non-Kozai plutino L4h15 (2004~HB$_{79}$; center),
and a non-resonant TNO (2004~PA$_{112}$; right) for comparison.
In both resonant cases, the resonant argument $\phi_{32}$ librates around 180$\degree$
(lower panels) because of the
3:2 mean-motion resonance with Neptune.
The Kozai plutino 28978 Ixion's integration also shows oscillations in $\omega$ around 270$\degree$ with coupled,
anti-correlated oscillations in $e$ and $i$.  
The integrations of the non-Kozai plutino and non-resonant TNO both show oscillations in $e$ and $i$, but they are not coupled,
and $\omega$ circulates.
Note that oscillations in semimajor axis and $\phi_{32}$ happen
on much faster timescales (few thousand years) than the Kozai oscillations (few million years),
and that the oscillations in semimajor axis are much larger for the resonant TNOs than the
non-resonant one.
}
\label{fig:orbint}
\end{figure*}

Resonances are diagnosed by inspecting the evolution of an object's orbital
elements during a numerical integration (Figure~\ref{fig:orbint}).
The integration must be long enough that the Myr-timescale Kozai oscillations
are visible.
If the object inhabits the $j$:$k$ mean-motion resonance with Neptune,
the primary resonant angle 
\begin{equation} \label{eq:generalres}
\phi_{jk}=j\lambda-k\lambda_N-(j-k)\varpi
\end{equation} 
will be confined and will not take on all values 0$\degree$ to 360$\degree$
over the course of the integration.
$\lambda$ is the mean longitude of the object 
\begin{equation} \label{eq:meanlongitude}
\lambda=\Omega+\omega+\curlym,
\end{equation}
which gives the angle to the position of the object relative to the reference direction.
$\lambda_N$ is the mean longitude of Neptune, giving the angle to
Neptune's position relative to the reference direction.
$\varpi=\Omega+\omega$ is the longitude of pericenter, which is the 
broken angle locating the object's pericenter relative to the reference direction.
The amplitude of the $\phi_{jk}$ oscillation during the integration
gives the libration amplitude $A_{jk}$ of the object.

All plutinos by definition inhabit the 3:2 mean-motion resonance with Neptune,
with a resonant argument 
\begin{equation} \label{eq:plutino}
\phi_{32}=3\lambda-2\lambda_N-\varpi.
\end{equation}

A feature of the Kozai resonance (discussed in detail in Section~\ref{sec:kozplutinos}) 
is coupled oscillation in $\omega$, eccentricity $e$, 
and inclination $i$ (see Figure~\ref{fig:orbint}).
$e$ and $i$ are anti-correlated, and $\omega$ oscillates around 90 or 270$^{\circ}$ 
\citep[or temporarily around 0 or 180$^{\circ}$;][]{NesvornyRoig2000,LykawkaMukai2007}.
Figure~\ref{fig:orbint} shows a 10~Myr orbital integration of the Kozai plutino 28978 Ixion, 
showing the characteristic oscillations of $e$, $i$, and $\omega$.
For comparison, Figure~\ref{fig:orbint} also shows a non-Kozai plutino and a non-resonant TNO
nearby in semimajor axis.

\section{Non-Kozai Plutinos} \label{sec:nonkoz}

To orient the reader and make several important points,
we first discuss toy models and then a realistic libration amplitude distribution for 
the plutinos, ignoring the Kozai component until Section~\ref{sec:kozplutinos}.

\subsection{0$\degree$ libration amplitude toy model} \label{sec:pluttoymodel0}

\begin{figure}
\centering
\includegraphics[scale=0.4]{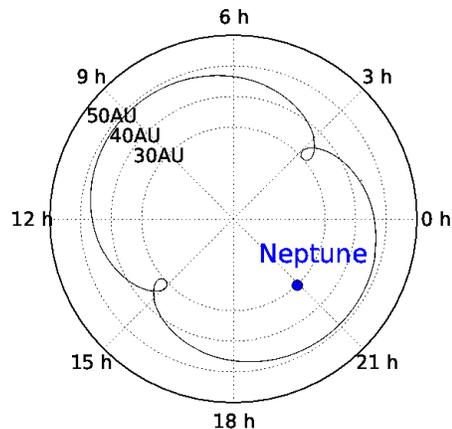}
\caption{The orbit of a 0$\degree$ libration amplitude, 0$\degree$ inclination, $e=0.24$ plutino in a frame that co-rotates with Neptune.
The motion in the co-rotating frame is clockwise, except at perihelion, where this example's
$e$ is so high that it briefly moves counterclockwise.
Pericenter occurs 90$\degree$ ahead and behind Neptune's position. 
Dotted circles are heliocentric distances of 30, 40, and 50~AU.
Neptune's position is shown for June 1, 2004 (which was midway through the CFEPS survey);
this is true for all subsequent plots.
}
\label{fig:simpleplut}
\end{figure}

Due to the plutino resonance condition (Equation~\ref{eq:plutino}), the location where the Plutinos can come to pericenter is restricted.
This is what allows resonant TNOs to remain stable on timescales of the age of the solar system,
despite having orbits that in some cases cross
the orbit of Neptune.
The resonant angle $\phi_{32}$ librates around 180$\degree$~$\pm$~360$\degree \times k$,
where $k$ is an integer (using $\phi_{32}$~=~180$\degree$ or -180$\degree$ is sufficient to
include plutinos at all angles from Neptune).  
At pericenter, $\curlym=0$, and at that moment equation~\ref{eq:meanlongitude} implies
$\lambda=\Omega+\omega=\varpi$.
For a plutino with libration amplitude $A_{32}=0\degree$, $\phi_{32}=180\degree$ always, and
\[180\degree=3\varpi-2\lambda_N-\varpi\]
\begin{equation} \label{eq:90lambdaN}
\varpi=\lambda_N+90\degree
\end{equation}
So the plutino always comes to pericenter 90$\degree$ away from Neptune's position.  
This is shown in Figure~\ref{fig:simpleplut}.
Because $\phi_{32}=-180\degree$ is also valid, another perihelion occurs
with $\varpi=\lambda_N-90\degree$.
The two points on the sky where a $A_{32}=0\degree$ plutino comes to pericenter
(in the ecliptic plane, at $\lambda_N\pm90\degree$),
are very important in this paper, so to avoid confusion we will refer to these
as the ``orthoneptune points''.

\subsection{95$\degree$ libration amplitude toy model} \label{sec:pluttoymodel95}

\begin{figure}
\centering
\includegraphics[scale=0.4]{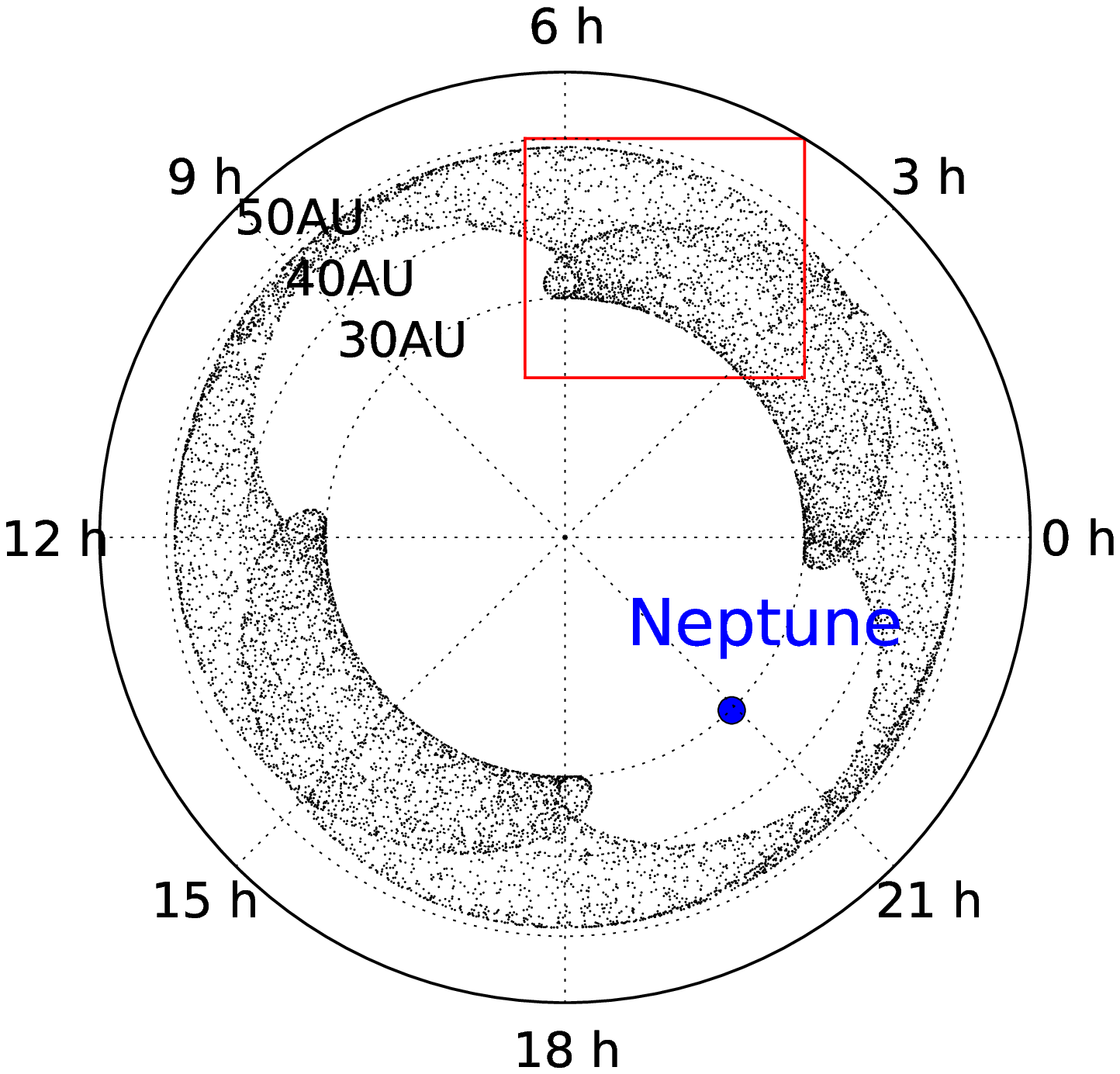} 
\includegraphics[scale=0.4]{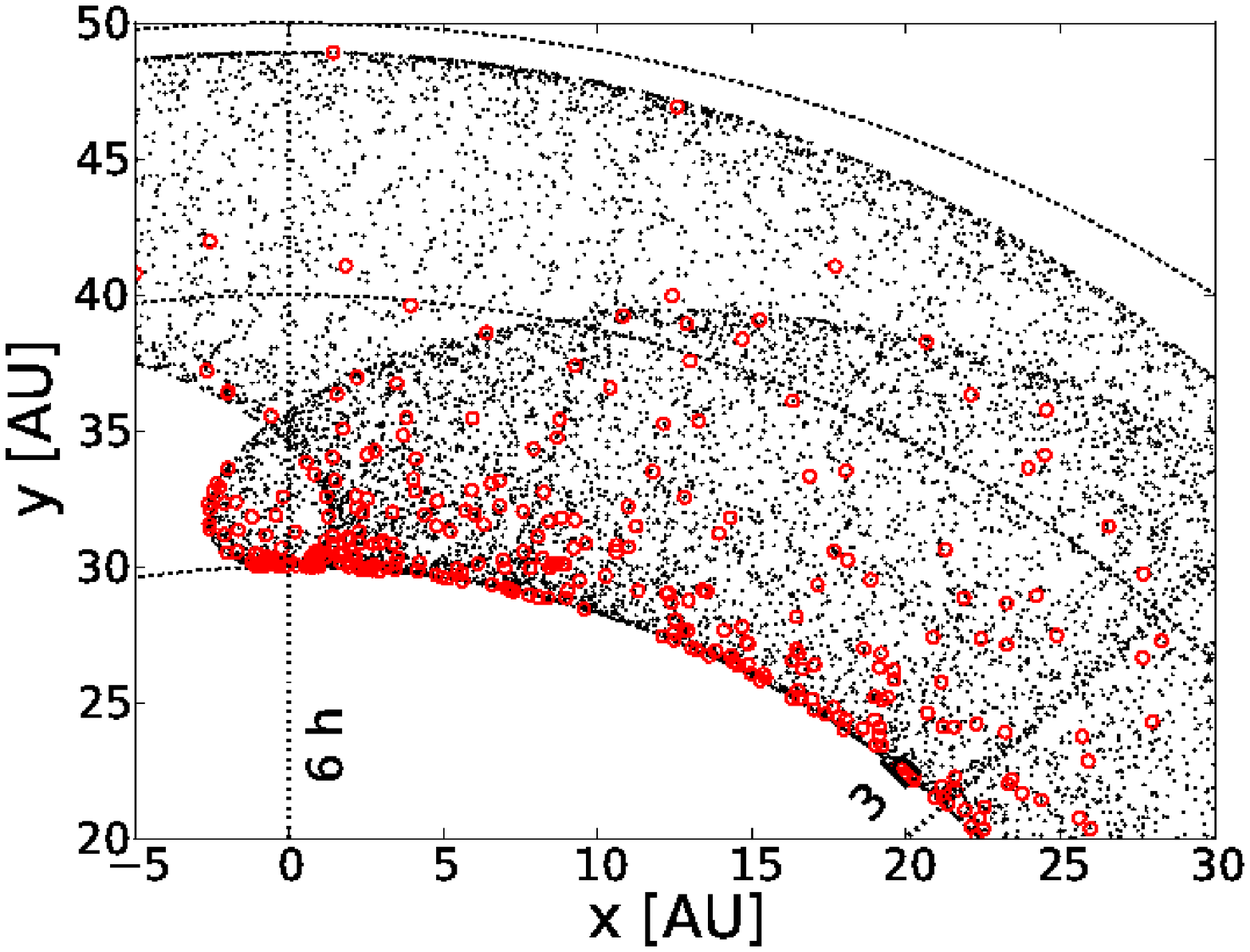}
\includegraphics[scale=0.4]{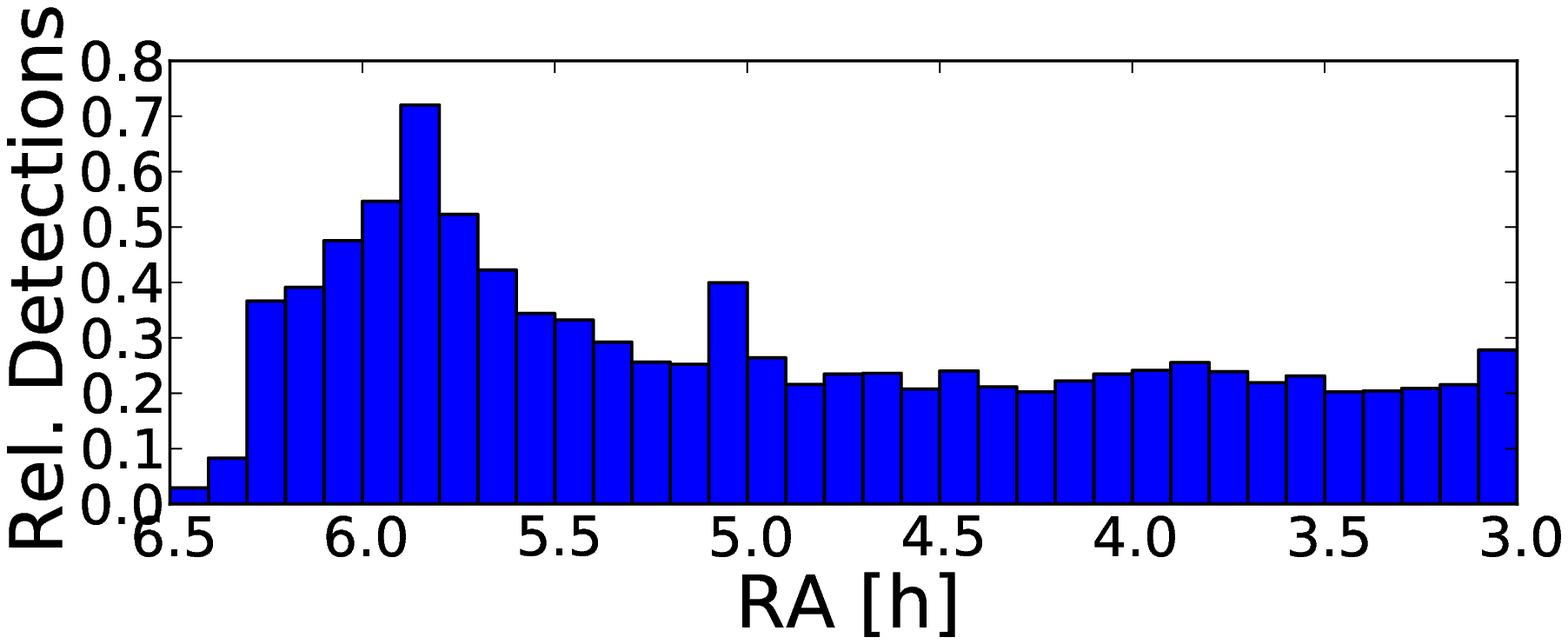}
\caption{Top panel shows a population of plutinos with $A_{32}=95\degree$, $i=0\degree$, and $e=0.24$
(see caption for Figure~\ref{fig:simpleplut}).
The 95$\degree$ libration amplitude means that the perihelion turnaround points correspond to orbits
with perihelia 42.5$\degree$ and 137.5$\degree$ ahead and behind Neptune.  
The red box is shown at higher resolution in the middle panel, where
red circles show simulated detections from a flux-limited all-sky survey.  
The lower panel shows that there are more detections per degree of ecliptic at the `turnaround' point,
which is where the libration causes the objects to come to pericenter farthest from the
orthoneptune points.
(The number of detections per bin is arbitrary.)}
\label{fig:turnaround}
\end{figure}

Real plutinos possess non-zero libration amplitudes; 
$A_{32}$ for known plutinos with well-characterized orbits ranges between 
20$\degree$ and 130$\degree$
\citep{LykawkaMukai2007, GladmanReson}.  
These libration amplitudes lead to different selection effects: 
during each $\phi_{32}$ libration period the perihelion direction 
oscillates around the orthoneptune points 
roughly sinusoidally in time with amplitude $A_{32}/2$.  
Equation~\ref{eq:plutino} shows that the maximum excursion
from the orthoneptune points occurs when $\phi_{32}$ is at a maximum or minimum.
This means that plutinos spend more time near the extrema allowed by their libration
amplitudes, and are actually more likely to be detected there.
We demonstrate this using a population of plutinos with $A_{32}=95\degree$ 
(see Figure~\ref{fig:turnaround} and caption).
To avoid confusion, all the plutinos in this toy model have 0$\degree$ inclination
to the ecliptic, and all have the same eccentricity.
$A_{32}=95\degree$ is chosen because it was found to be the most common plutino libration
amplitude by CFEPS \citep{GladmanReson}.

\subsection{A realistic plutino distribution} \label{sec:plutrealdist}

\begin{figure}
\centering
\includegraphics[scale=0.4]{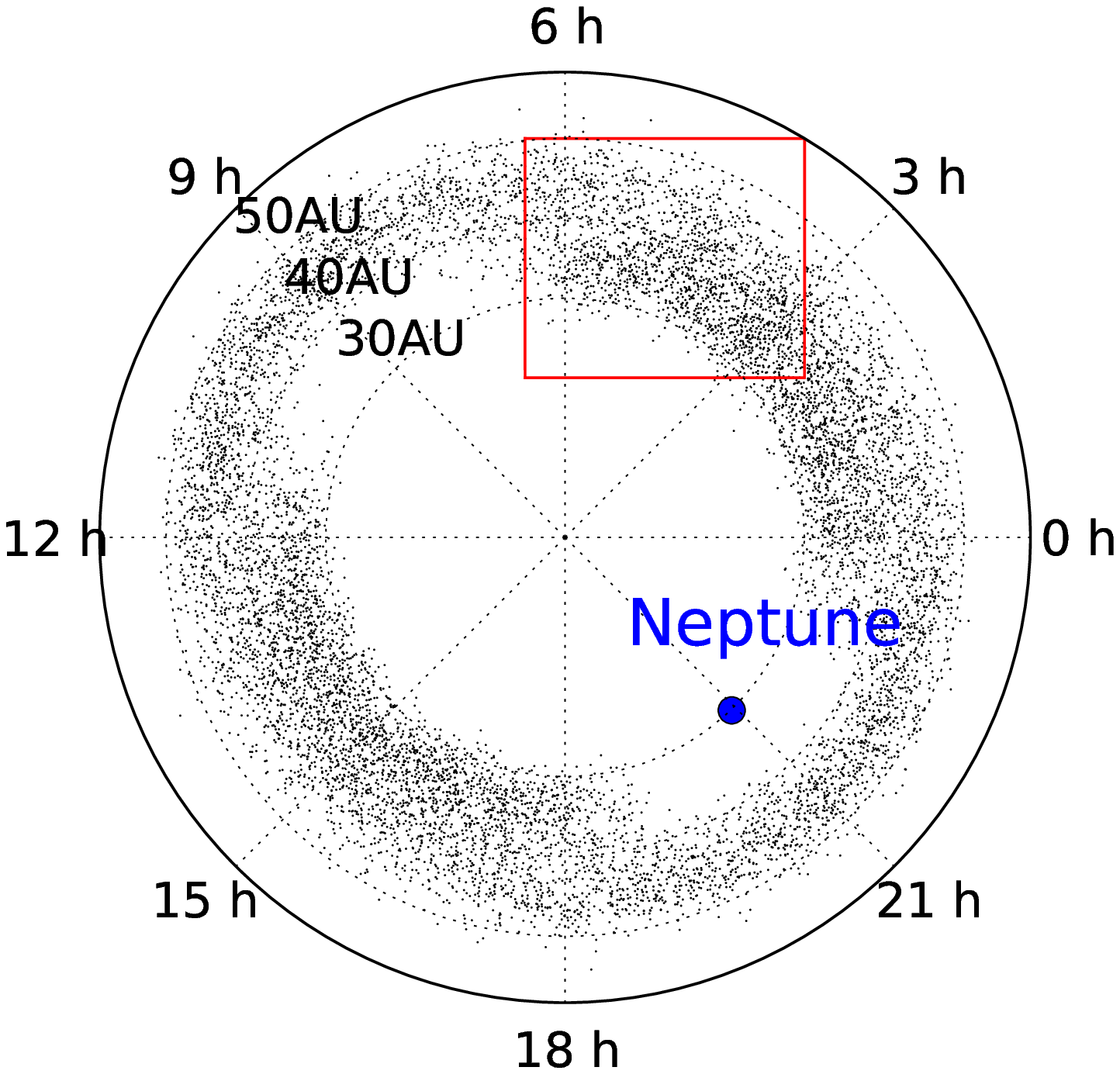} 
\includegraphics[scale=0.4]{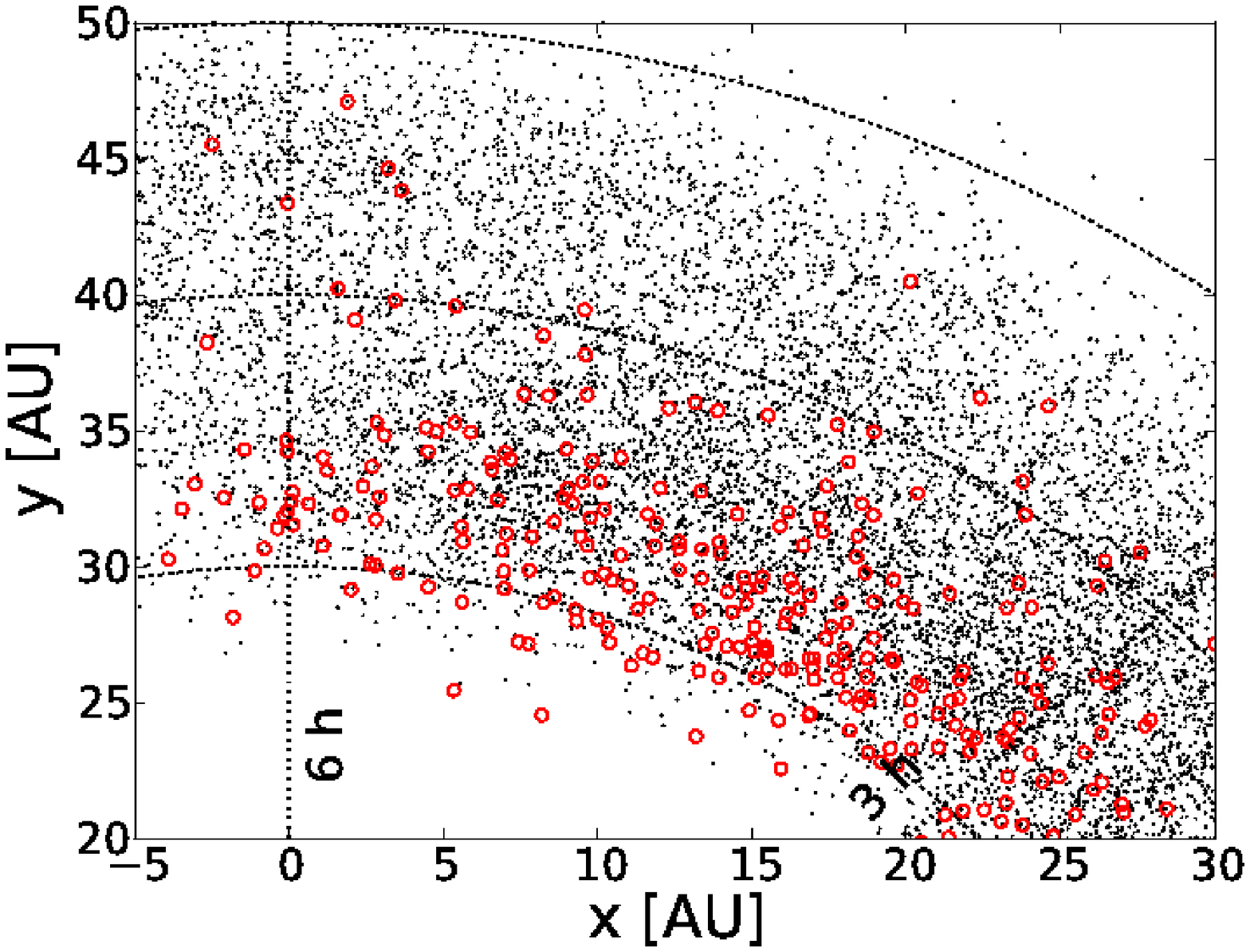}
\includegraphics[scale=0.4]{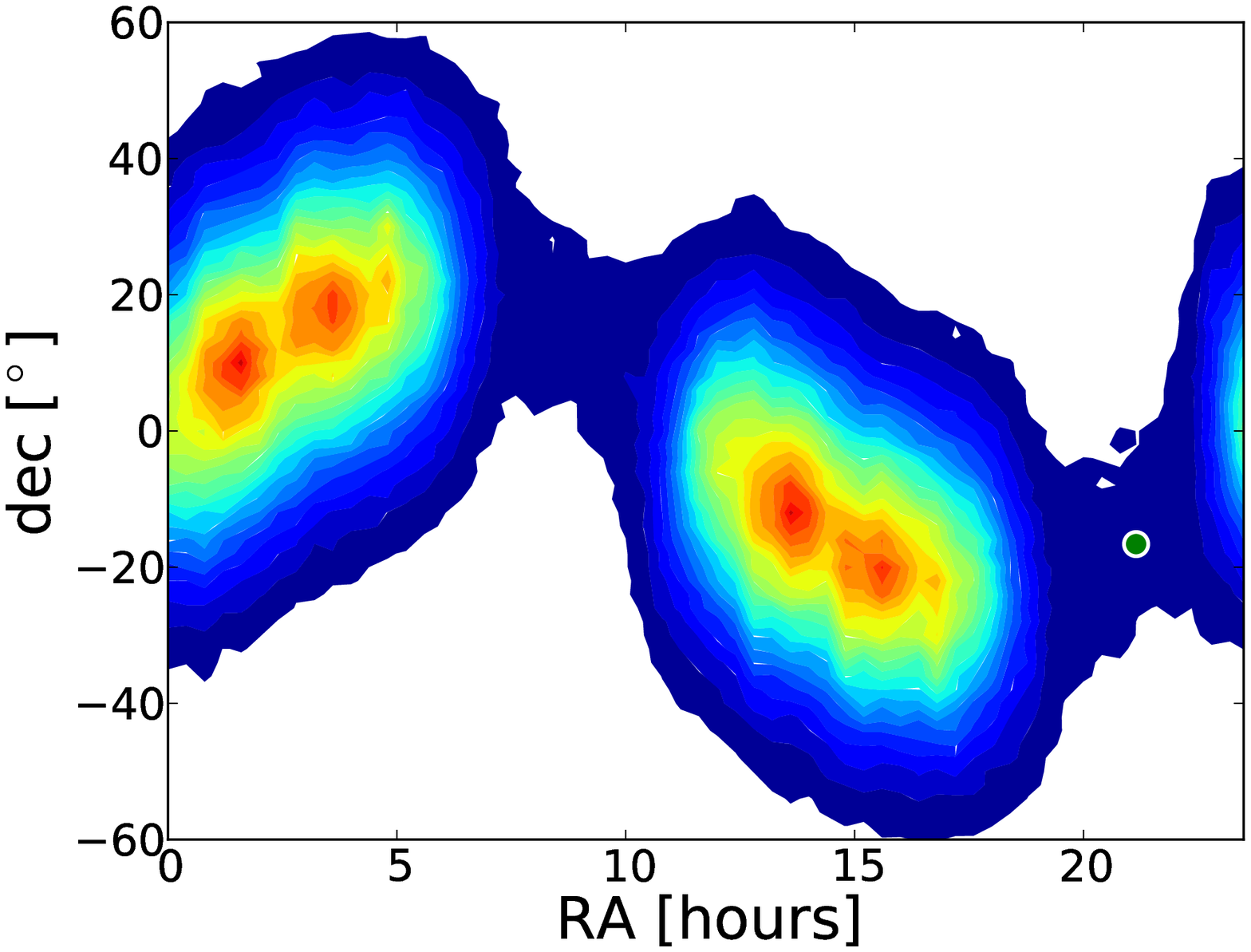}
\caption{Top panel shows a population of plutinos with a distribution of orbital 
elements that matches the debiased
plutino model from CFEPS \citep[][excluding the Kozai component]{GladmanReson}.  
See Figure~\ref{fig:turnaround} caption.
The lower panel here shows the detection density on the sky, 
with red being higher density and blue being lower, with contours evenly spaced
in detection density.
Two peaks are visible on either side of the orthoneptune points, 
which are caused by the turnaround effect.
At these peaks, the detection density is $\sim$30\% higher than at the
orthoneptune points.
}
\label{fig:turnaroundfuzzy}
\end{figure}

In actuality, plutinos possess a range of libration amplitudes.  
Figure~\ref{fig:turnaroundfuzzy} shows an observationally-motivated 
distribution of plutinos, 
based on the debiased model from CFEPS \citep[][excluding the Kozai component, which is discussed
in detail in section~\ref{sec:kozplutinos}]{GladmanReson}. 
\citet{ChiangJordan2002} and \citet{Malhotra1996} also presented models of the plutino
distribution.
\citet{Malhotra1996} discusses the dynamics of plutinos for given values of the libration amplitude,
while \citet{ChiangJordan2002} examined distributions of particles where $A_{32}$ was established in a cosmogonic simulation.
In contrast, our distributions of $a$, $e$, $i$, and $\phi_{32}$ are determined by 
debiasing the CFEPS Survey.

While at first glance it appears that the `turnaround' effect shown in 
Figure~\ref{fig:turnaround} 
is completely lost, this is not the case.  
Each plutino is still most likely to be detected at its maximum perihelion excursion 
from the orthoneptune points of $A_{32}/2$.
So, a plutino with an 80$\degree$ libration amplitude is most likely to be detected 
40$\degree$ away from the orthoneptune points, at 
$\lambda_N\pm50\degree$ or $\lambda_N\pm130\degree$, 
while a plutino with a 20$\degree$ libration amplitude is most likely to be detected 
10$\degree$ away from the orthoneptune points, at 
$\lambda_N\pm80\degree$ or $\lambda_N\pm100\degree$.

Because CFEPS showed that plutinos with $A_{32}<20\degree$ are so rare as to be 
approximated as absent,
two peaks in the detectability are visible, about 15$\degree$ on either side of 
the orthoneptune points.
As Figure~\ref{fig:turnaroundfuzzy} shows, this means that the place on the sky
to find the most plutinos is not 90$\degree$ away from Neptune.
For the CFEPS L7 model of the true libration amplitude distribution
\citep{GladmanReson}, 
the maximum on-sky detection rate 
(integrated over all libration amplitudes) happens
about $\pm15\degree$ away from the orthoneptune points.

\section{Kozai Plutino Dynamics} \label{sec:kozplutinos}

\begin{figure*}
\centering
\includegraphics[scale=0.38]{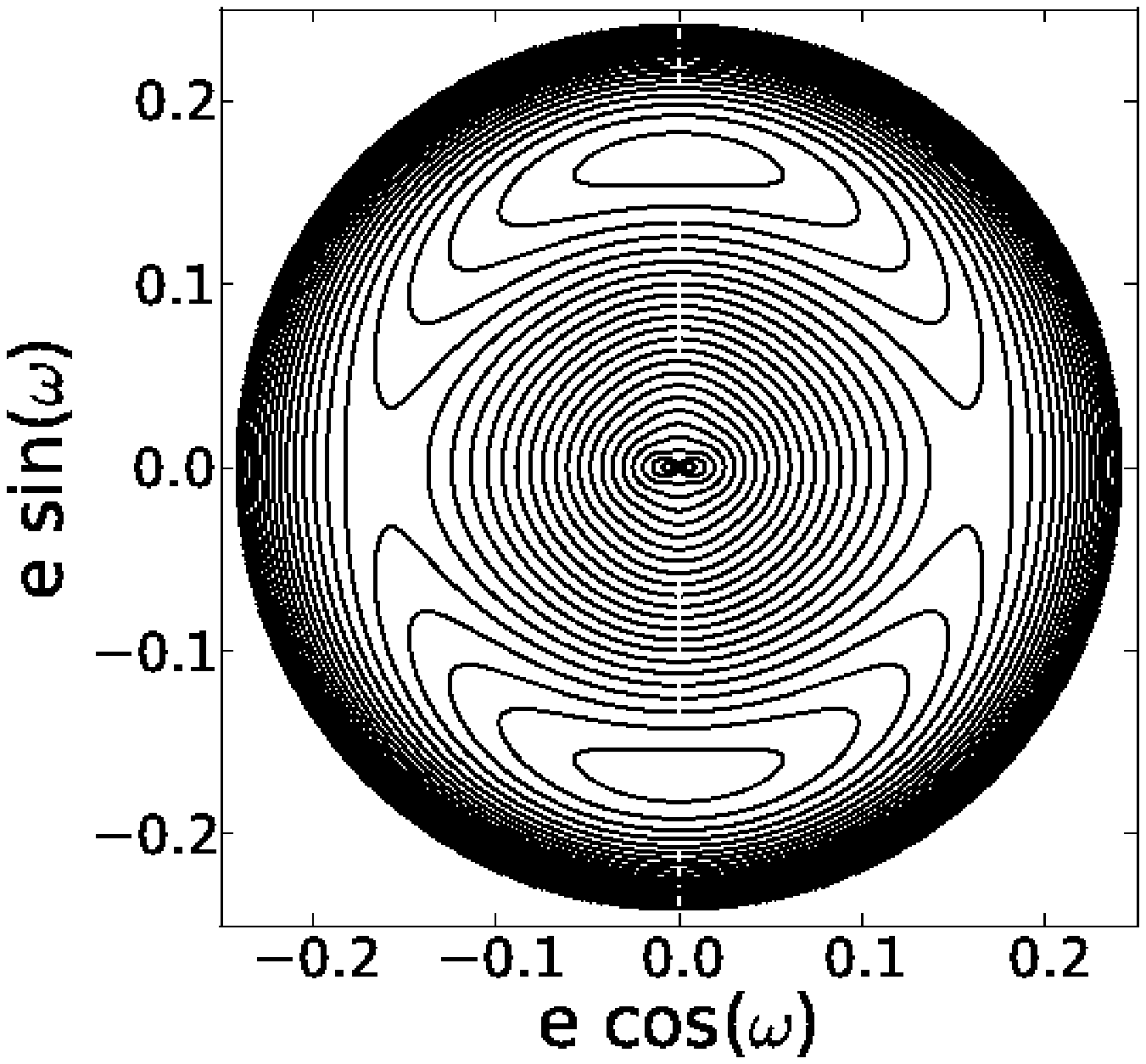} \includegraphics[scale=0.38]{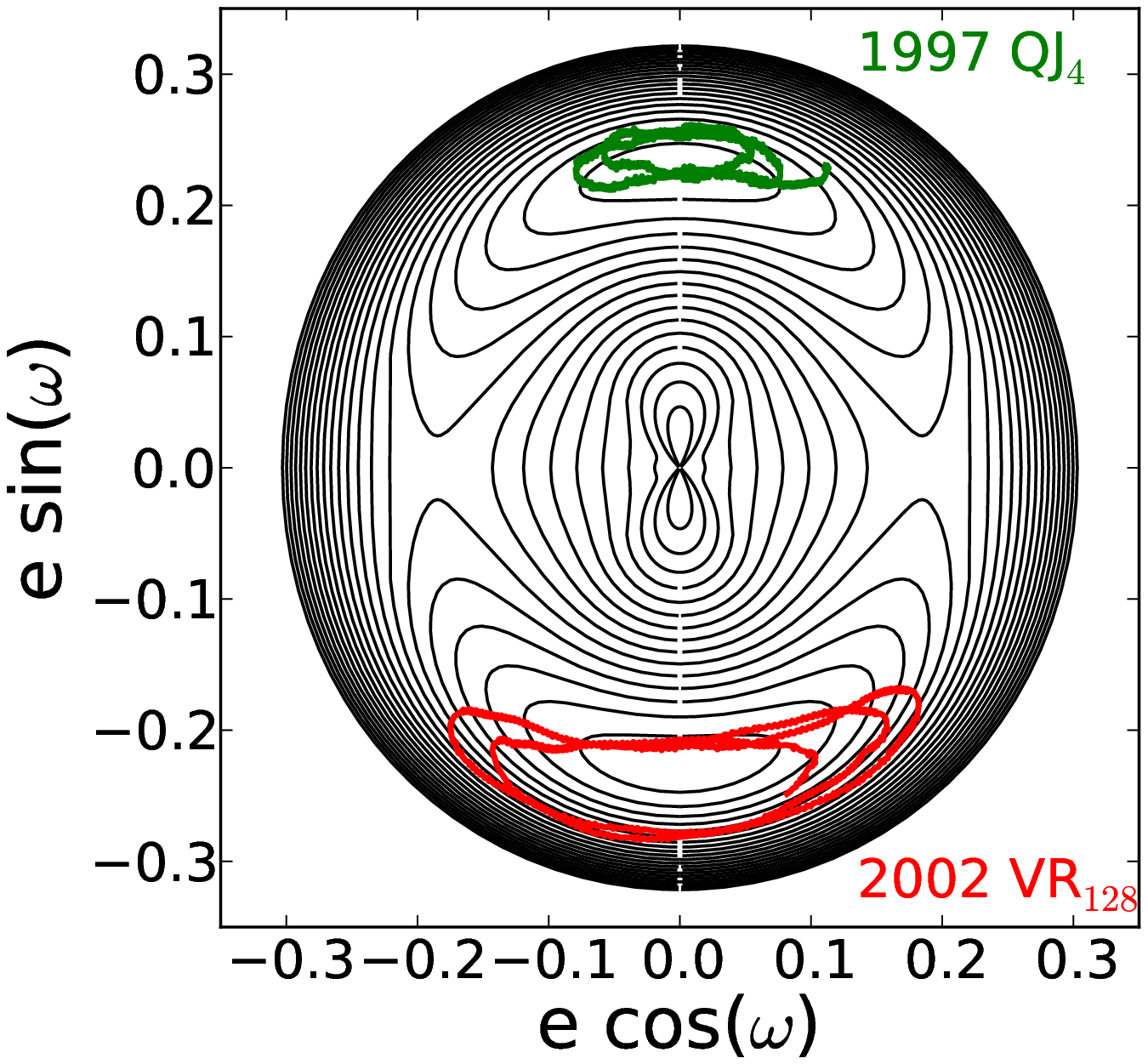} \includegraphics[scale=0.38]{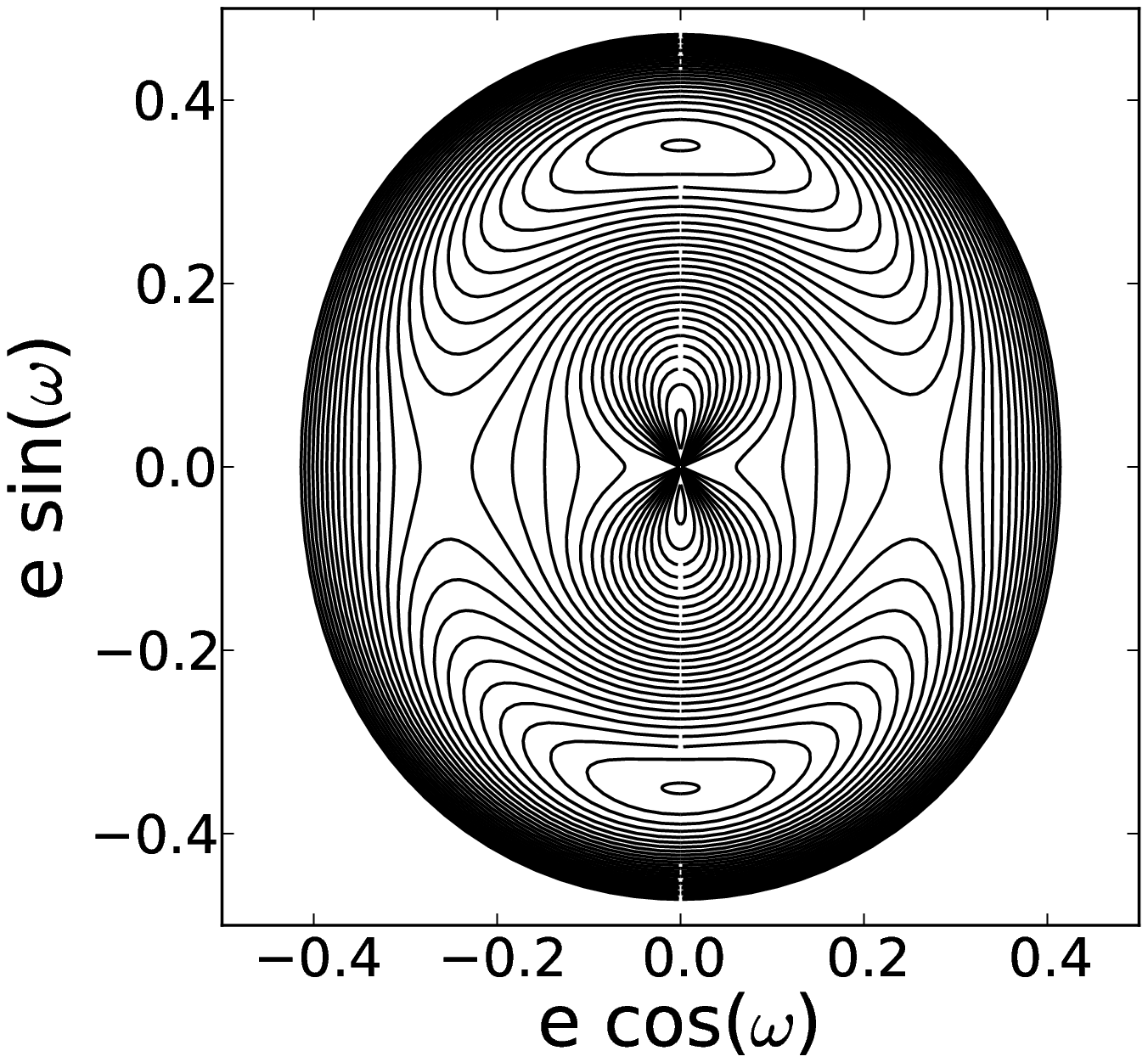}
\caption{3 different Hamiltonian level surfaces for plutinos 
\citep[constructed using the disturbing function from][]{WanHuang2007},
for different values of angular momentum.
Left panel is $\imax=14\degree$, center is $\imax=21\degree$, and right is $\imax=34\degree$.  
These are polar plots, with $e$ as the radius and $\omega$ as the angle.
Also shown on the center plot are the 10~Myr orbital integrations of the Kozai plutinos 1997~QJ$_4$
and 2002~VR$_{128}$, showing circulation around the $\omega=90\degree$ or 270$\degree$ islands over time;
these plutinos were chosen because they have $\imax\simeq21\degree$.
While the presence of other planets causes small changes in the Hamiltonian, one can see
that the evolution is decently approximated.
}
\label{fig:kozham}
\end{figure*}

This section discusses the dynamics of objects that are simultaneously in the 3:2 
mean-motion resonance with Neptune (plutinos) and in the Kozai resonance, and
the effects these two simultaneous resonances have on the on-sky detectability.

The Kozai resonance can occur at much 
lower inclinations within mean-motion resonances than for non-resonant TNOs
\citep{ThomasMorbidelli1996, WanHuang2007}.
The $\omega$ oscillation is unique to the Kozai resonance: the perturbations of 
the other solar system planets on a small body (resonant or not) normally cause
$\omega$ to precess rather than librate.

The libration in $e$, $i$, and $\omega$ can be best understood using a contour plot
of the averaged Hamiltonian of the disturbing function, which describes 
secular oscillation due to the three-body interaction.  
Example surfaces for the 4th order disturbing function for a Kozai plutino \citep{WanHuang2007}
are shown in Figure~\ref{fig:kozham}.
These are polar plots, where the length of the vector gives $e$, 
and the angle from 0$\degree$ gives $\omega$.
In the cases shown, 
only the contours that close around 90$\degree$ or 270$\degree$ correspond to 
Kozai oscillations.
Each plutino that is also in the Kozai resonance is confined to a particular contour 
on one of these surfaces.
Tracing one contour reveals how $e$ and $\omega$ vary during the course
of a Kozai cycle, with the range in $\omega$ values describing
the Kozai libration amplitude $A_{\omega}$ around the relevant libration center.
Orbital inclination is calculated using $e$ and conservation of the z-component of angular
momentum $L_z~\propto~{\rm cos}~i\sqrt{1-e^2}$, because
\begin{equation} \label{eq:Lz}
{\rm cos}~i\sqrt{1-e^2}={\rm cos}~i_0\sqrt{1-e_0^2}
\end{equation}
for initial inclination $i_0$ and initial eccentricity $e_0$ at any time.
Each surface plot has its own $L_z$ value, which is parameterized using $\imax$:
\[{\rm cos}~i_0\sqrt{1-e_0^2}={\rm cos}~\imax\]
$\imax$ is the inclination required by conservation of angular momentum for $e$=0.
Note that $\imax$ is not the maximum inclination that these plutinos will reach; 
in order for an object to have that inclination, 
it needs $e$=0, which will not happen in the course of a high-$e$ Kozai oscillation.
Plutinos always have maximum inclination values attained during their Kozai
oscillations that are less than $\imax$.
$\imax$ is just a way to parameterize the level surfaces.

For a known Kozai plutino, the level surface can be chosen
using the measured $e_0$ and $i_0$ values to calculate $\imax$, which
gives the Hamiltonian level surface for this object.
The measured $\omega_0$ value sets which contour the object is oscillating on,
and knowing the contour allows the Kozai libration amplitude 
$A_{\omega}$ to be calculated numerically.

\subsection{On-Sky Detection Biases for Kozai Plutinos} \label{sec:kozonsky}

A direct consequence of the Kozai resonance-caused oscillation of $\omega$ is that these 
objects always come to pericenter out of the ecliptic plane.  
Because these are plutinos, we start with the same resonant condition (equation~\ref{eq:plutino}), 
and for this illustration choose $A_{32}=0\degree$:
\[\phi_{32}=3\lambda-2\lambda_N-\varpi\]
\[180\degree=3(\Omega+\omega+\curlym)-2\lambda_N-(\Omega+\omega)\]
One can see from Figure~\ref{fig:kozham} that when the Kozai libration
amplitude is 0$\degree$, $\omega$=90$\degree$ or 270$\degree$.
At pericenter, $\curlym=0\degree$.
Substituting these values into the above equation yields $\Omega=\lambda_N$,
indicating that the node of the plutino orbit is in the same direction as Neptune, 
and $\omega$ will be $90\degree$ ahead or behind that position 
($\Omega=\lambda_N+180\degree$ is also valid).
The orbital plane of the plutino will be tilted out of the ecliptic plane by the inclination,
with the line of nodes through Neptune's position acting as the pivot.
Most of the Kozai plutinos listed in the MPC database have orbital 
inclinations between 10$\degree$ and 30$\degree$ 
\citep{LykawkaMukai2007}, which means that when they are at pericenter, a 0$\degree$ Kozai libration
amplitude plutino will be roughly $10-30\degree$ above 
or below the ecliptic plane.

\begin{figure}
\centering
\includegraphics[scale=0.4]{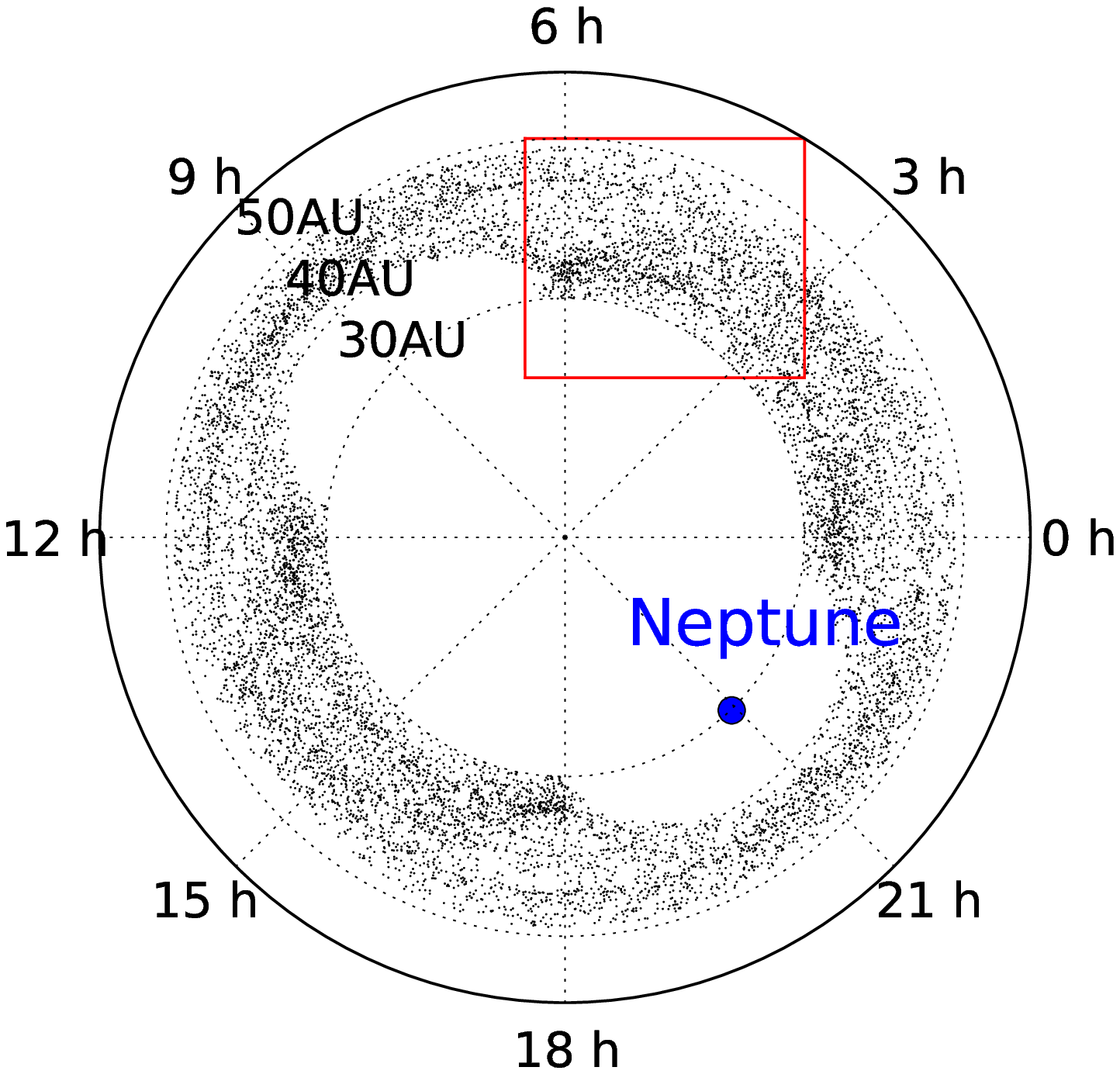} 
\includegraphics[scale=0.4]{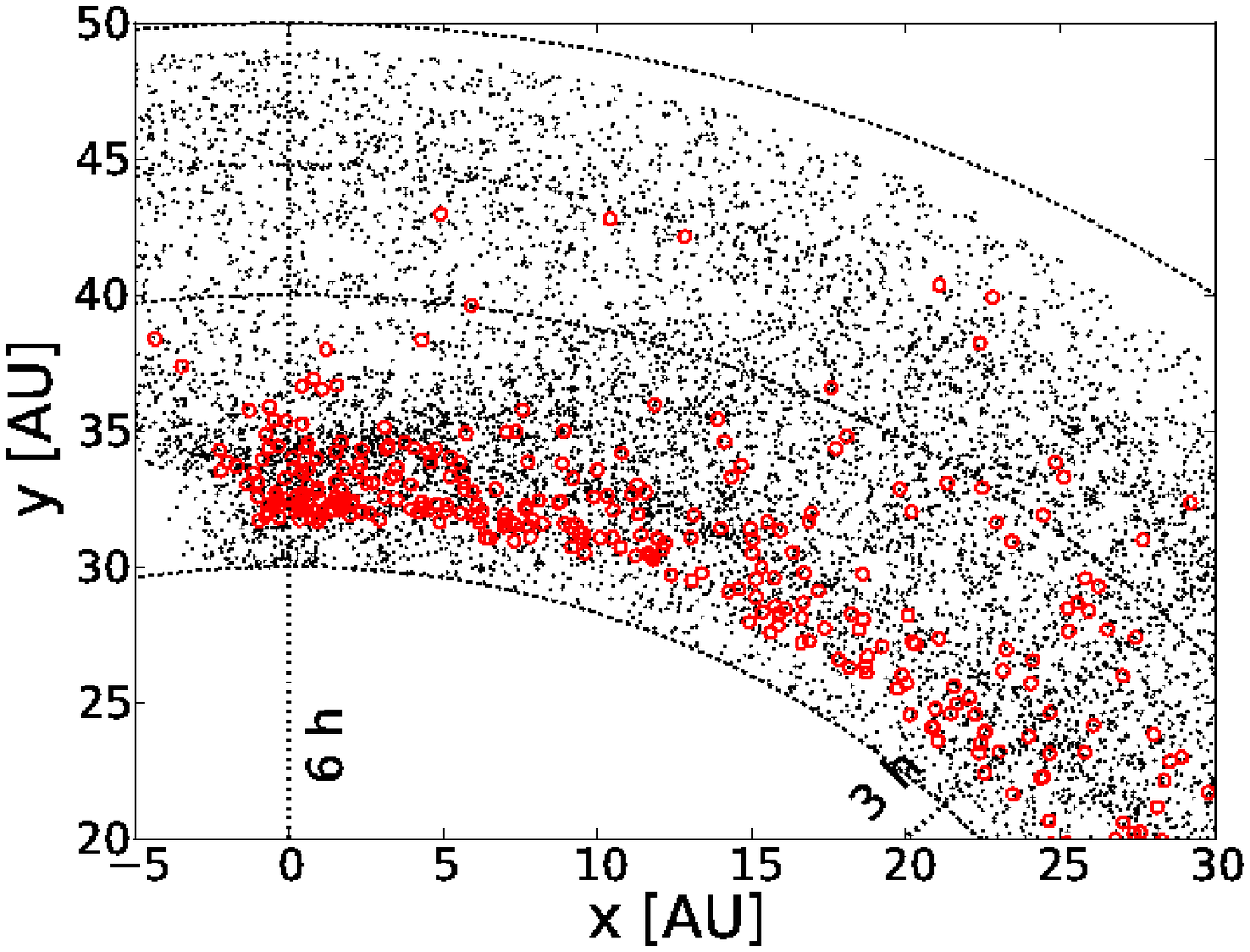}
\caption{Top panel shows a toy model of the Kozai plutinos, where all objects have $A_{32}=95\degree$,
using one contour from one level surface (meaning that all these Kozai plutinos have the same Kozai
libration amplitude $A_{\omega}$).
Objects in the red box are shown at higher resolution in bottom panel.
Red circles show simulated detections from a flux-limited ecliptic survey.  
The objects are only detected where they cross through the ecliptic plane, which is never
when the objects are at pericenter.
}
\label{fig:turnaround_kozai}
\end{figure}

\begin{figure}
\centering
\includegraphics[scale=0.4]{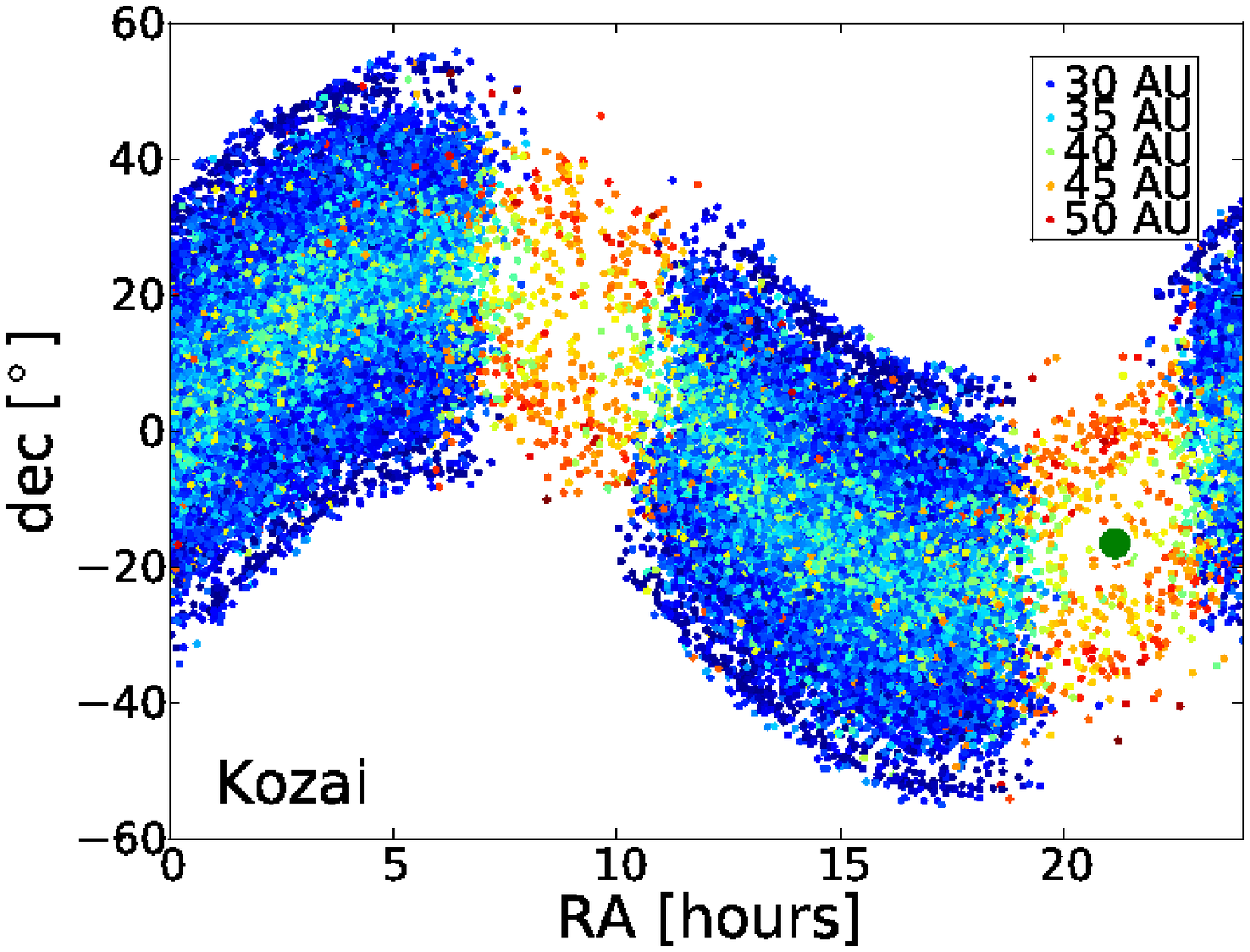} 
\includegraphics[scale=0.4]{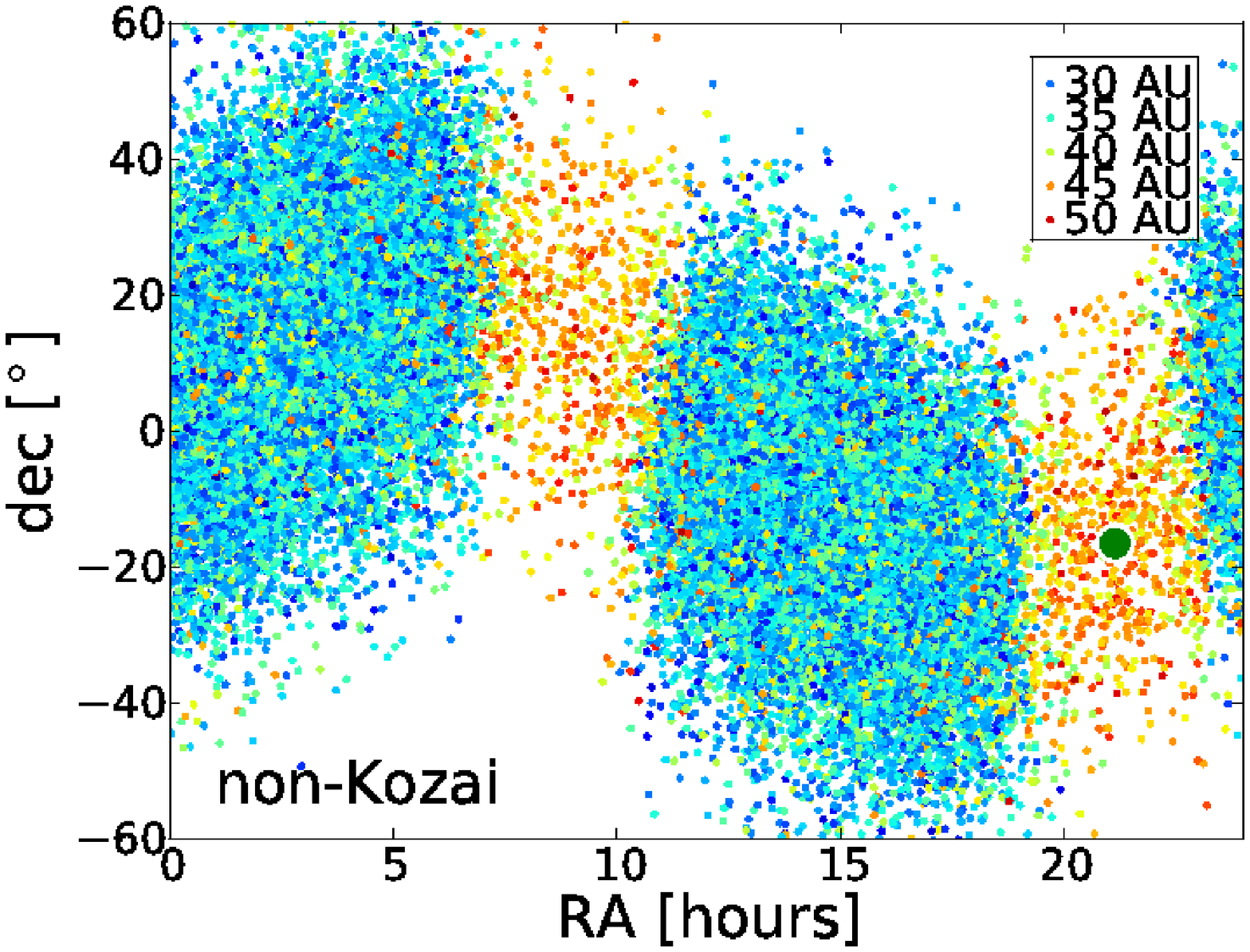}
\includegraphics[scale=0.4]{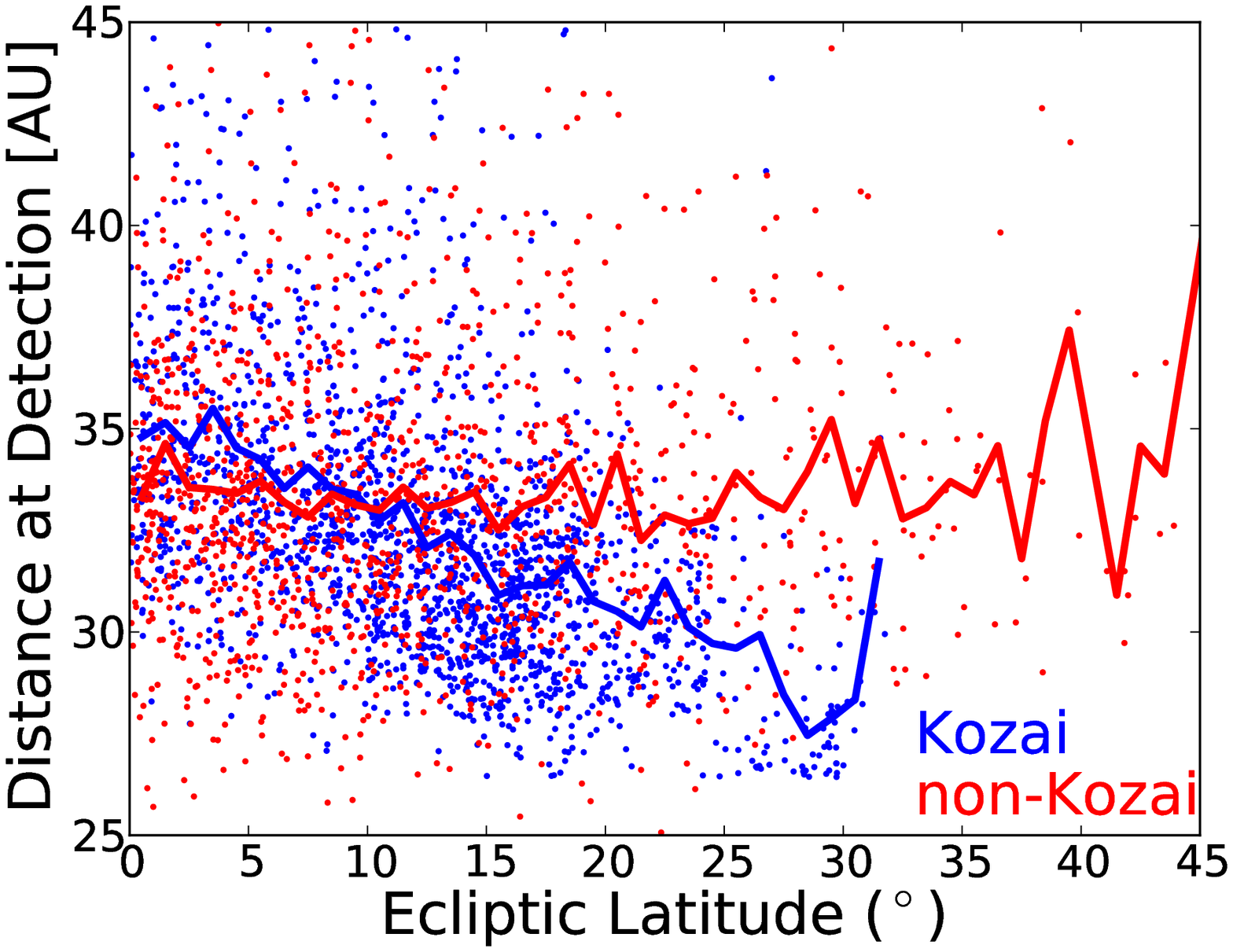}
\caption{
Detected plutinos in a simulated, flux-limited, all-sky survey,
using realistic distributions for both populations.
In the two upper panels, the color of each point shows the distance at which the plutino was 
detected (see legend in panels),
with Kozai plutinos in the uppermost panel and non-Kozai plutinos in the middle panel.
The non-Kozai plutinos are detected at a broad distance range in each of the two pericenter lobes.
The distance at which the Kozai plutinos are detected is very much dependent on ecliptic latitude,
with higher ecliptic latitudes detected at closer distances, and lower detected at greater distances.
The large green dot marks Neptune's location.
The lower panel shows the distance and ecliptic latitude where each plutino was detected, 
using one of the pericenter lobes (plutinos between ecliptic longitudes of 12h-18h).
Blue shows Kozai plutinos, red shows non-Kozai.  
The solid lines show the average in 0.5$\degree$ bins of ecliptic latitude.
The average detection distace for Kozai plutinos depends on ecliptic latitude,
with those at low latitudes being at the farthest average heliocentric distance.
Non-Kozai plutinos show no such trend.
}
\label{fig:distanceofdetection}
\end{figure}

Solar System objects are most easily detected at pericenter, when they are closest and thus brightest.
Because the Kozai plutinos are forced to be out of the ecliptic at pericenter, they will be harder
to detect in ecliptic surveys than non-Kozai plutinos 
(see Figures~\ref{fig:turnaround_kozai} and \ref{fig:distanceofdetection}).
This is an important bias that must be accounted for.

\section{Simulating the Plutinos} \label{sec:pop}

\begin{figure*}
\centering
\includegraphics[scale=0.28]{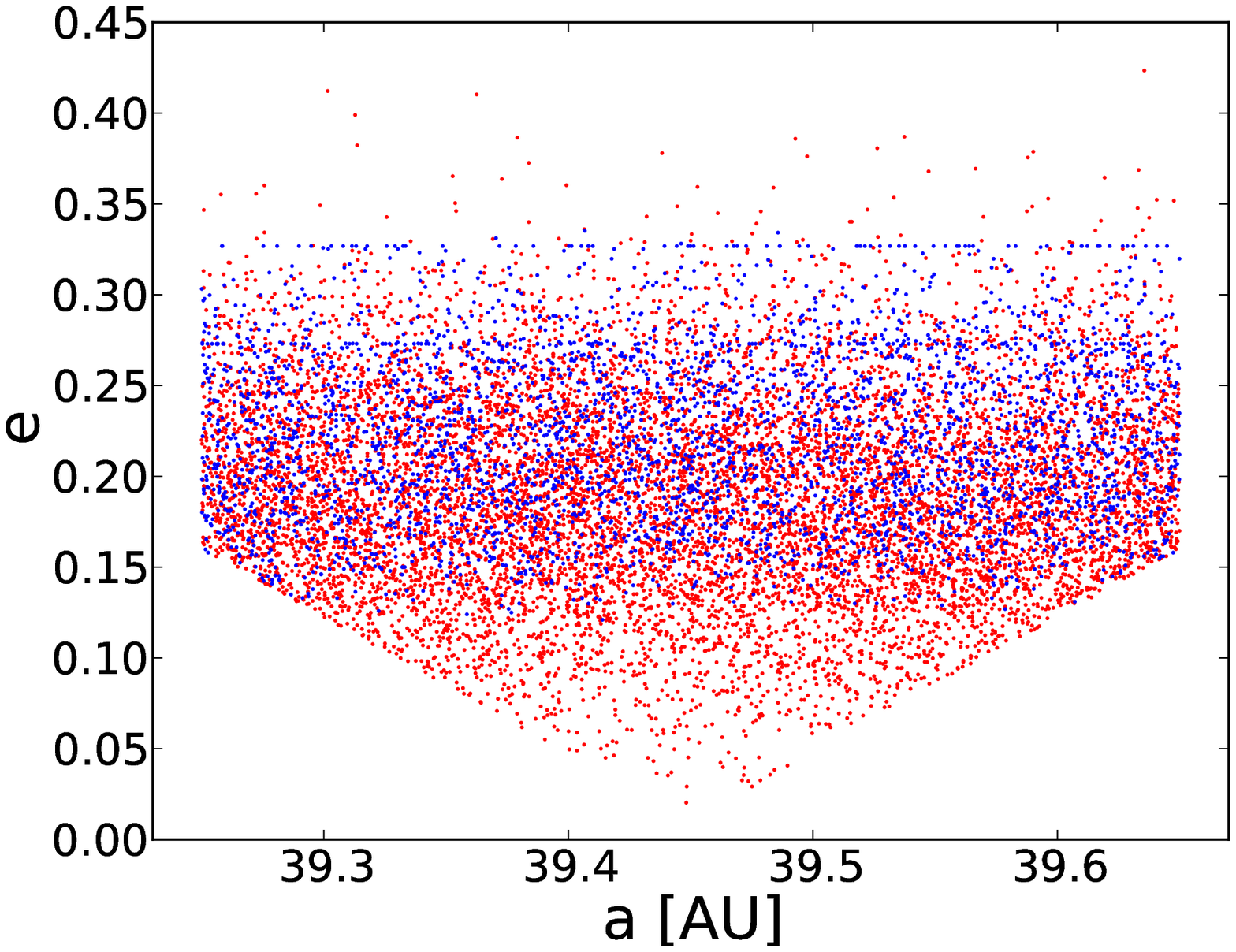} \includegraphics[scale=0.28]{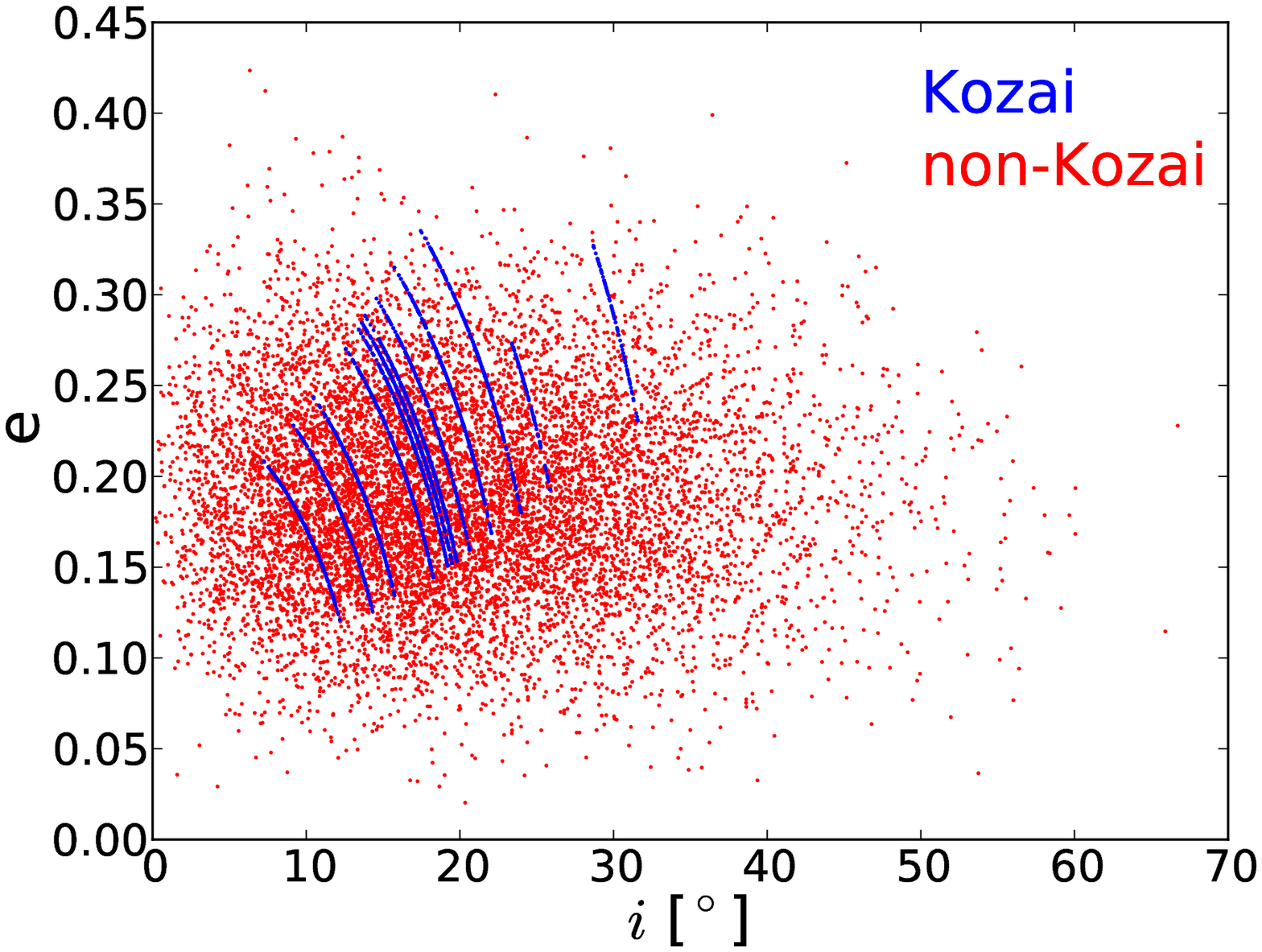} \includegraphics[scale=0.28]{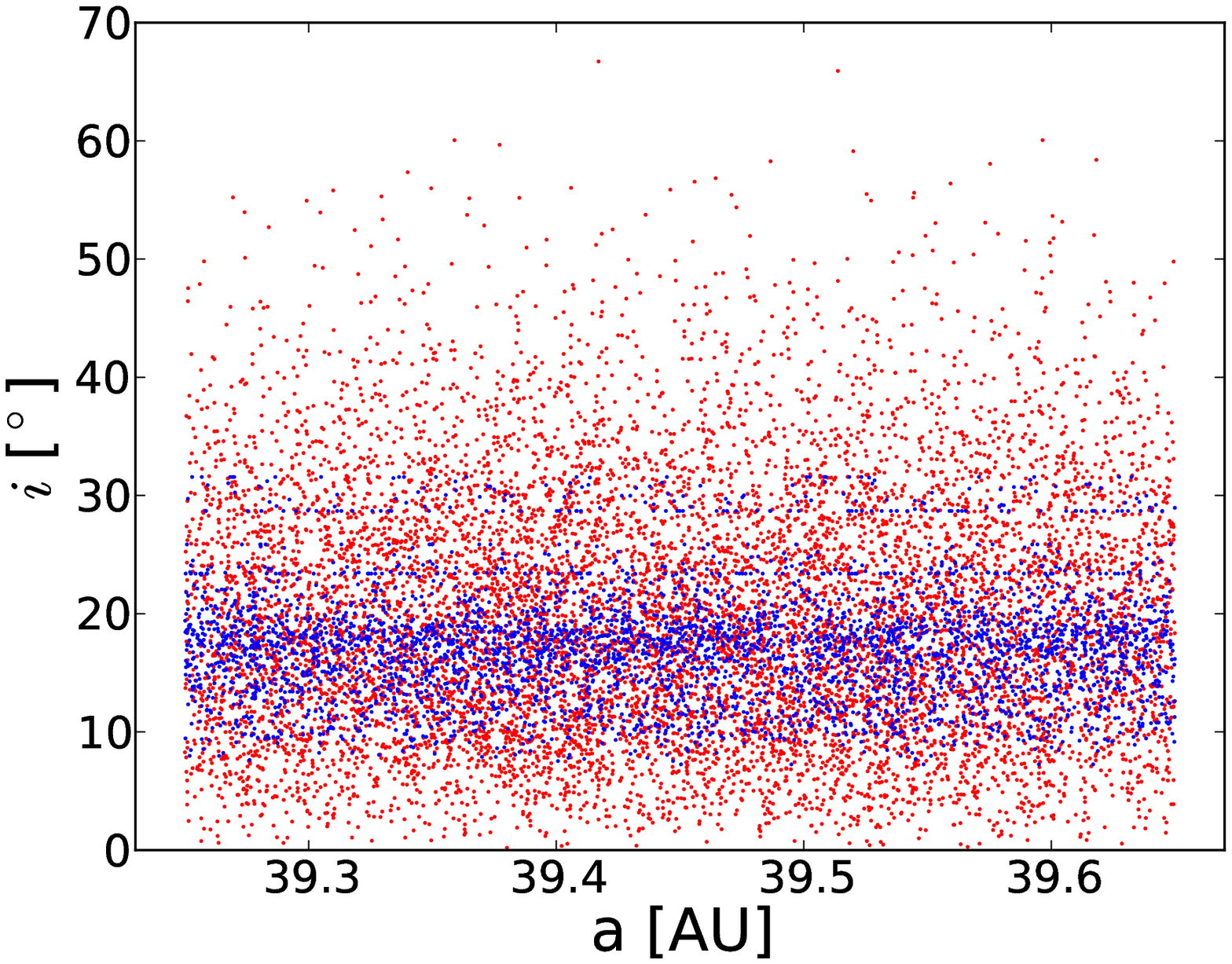}
\caption{Distribution of simulated plutinos in semimajor axis, eccentricity, and inclination.
Red points are non-Kozai plutinos and blue points are plutinos in the Kozai resonance.
The Kozai plutinos are constrained by the conservation of $L_z$ and our choice of $\imax$ values.
This is especially obvious in the middle panel, where each Kozai curve corresponds to a 
particular $\imax$ value;
over a Kozai libration cycle, objects never reach $e=0$, and thus always have $i<\imax$.
}
\label{fig:plutinoaei}
\end{figure*}

We simulate the population of plutinos by randomly drawing from specified distributions
of orbital elements and magnitudes, then taking each simulated plutino and ``observing'' it using
the CFEPS survey simulator (publicly available at {\tt www.cfeps.net}).
Figure~\ref{fig:plutinoaei} shows the simulated Plutino distribution in semimajor axis,
eccentricity, and inclination.
The non-Kozai plutino distribution we use here is identical to the CFEPS L7 model,
while the Kozai plutino distribution here is more detailed than that used by \citet{GladmanReson}.

The following describes how our code builds a population of plutinos 
with a realistic orbital element distribution, one simulated object at a time.
The first step is to choose whether or not a given object is in the Kozai resonance
or not.  

An important parameter in these simulations is the true Kozai fraction $\fkoztrue$,
which is the true number of Kozai plutinos divided by the total number of plutinos.
This is not necessarily the same as the observed Kozai fraction $\fkozobs$,
which is the total number of detected Kozai plutinos divided by the total number
of detected plutinos for a given survey.
For most of our simulations, we use a true Kozai fraction $\fkoztrue=10\%$, 
based on the results of CFEPS.
However, because \citet{GladmanReson} found that Kozai fractions up to 33\% cannot be ruled
out at the 99\% confidence level, some of our calculations are repeated for 
$\fkoztrue$ values of 20\% and 30\%.
Constraining the value of $\fkoztrue$ will require many more well-characterized plutino 
detections than are currently available.

\subsection{Non-Kozai Plutinos} \label{sec:pop:nonkoz}

For the plutinos which are {\it not} also in Kozai (with percentage 100\%-$\fkoztrue$),
the following 
procedure is followed to choose its orbital elements.
This is the same as the best-fit non-Kozai plutino model from CFEPS \citep{GladmanReson}.

First, the eccentricity is chosen from a Gaussian probability distribution
centered on 0.18 with a width of 0.06.
Eccentricities large enough to approach the orbit of Uranus ($e>0.22$) are not allowed.
The semi-major axis is chosen from a simplified version of the stability
tests of \citet{TiscarenoMalhotra2009}.
$a$ is chosen within 0.2~AU of 39.45~AU for objects with $e>0.15$.  
The allowed range of $a$ values drops linearly as $e$ gets smaller, reaching 
a width of zero at $e=0$ (see Figure~\ref{fig:plutinoaei}).
The inclination is then chosen from 
a probability distribution of the form 
\[
P(i) \propto \sin i \exp \left( \frac{-i^2}{2\sigma_{32}^2} \right)
\]
with $\sigma_{32}=16^\circ$ 
\citep[originally based on the inclination distribution postulated by][]{Brown2001}.
The libration amplitude is chosen from an asymmetric ``tent-shaped'' probability distribution 
with a peak at 95$^{\circ}$,
and linearly decreasing probabilities to the lower and upper bounds, 
20$^{\circ}$ and 130$^{\circ}$ respectively, where the probability drops to zero.

Lastly, the object's absolute magnitude $H_g$ is chosen from an 
exponential distribution:
$N(<H_g)~\propto~10^{\alpha H_g}$, with $\alpha$=0.9.  
The reader is cautioned that this $\alpha$ value can only be considered
valid in the range of $H_g$ magnitudes where CFEPS had many detections 
(approximately $8<H_g<9$ for the plutinos).
Sensitivity to the size distribution is 
discussed further in Section~\ref{sec:alpha}.

\subsection{Kozai Plutinos} \label{sec:pop:kozai}

For a Kozai plutino, a slightly different path is followed
to choose its orbital elements.

First, the Hamiltonian level surface is chosen.  
Inspecting the results of \citet{LykawkaMukai2007}, we found which Hamiltonian level 
surface corresponded to each of their Kozai plutinos.  
To reflect this distribution of surfaces, we used level surfaces corresponding to $\imax$
of 14$\degree$, 16$\degree$, 17.5$\degree$, 20$\degree$, 21$\degree$, 21.3$\degree$, 
21.6$\degree$, 22.5$\degree$, 24$\degree$, 26$\degree$, 28$\degree$, and 34$\degree$,
in equal proportions.
In reality, due to the historical dominance of ecliptic surveys and the bias against detecting 
large $i$ TNOs, there are probably a larger fraction of large $i$ Kozai librators;
however, the currently available information does not justify more complex modeling.

Once the $\imax$ level surface is chosen, we pick a Kozai libration amplitude $A_{\omega}$
at random between 10$\degree$-80$\degree$.
$\omega$ is then chosen sinusoidially within the values allowed by $A_{\omega}$.
Because of the banana-shape of the contours, there are two values of $e$ allowed for 
any given value of $\omega$ (see Figure~\ref{fig:kozham}).
Given this value of $\omega$, the \citet{WanHuang2007} disturbing function allows numerical
determination of the two $e$ values that correspond to $\omega$ on the contour.
Half of the time we choose the lower value of $e$, and half higher.
The inclination $i$ is calculated using ${\rm cos}~i\sqrt{1-e^2}={\rm cos}~\imax$.
Since this only covers the 90$^{\circ}$ Kozai libration island, half of the objects are 
flipped $\omega$ of to 360$^{\circ}$ minus the original $\omega$ value. 

Lastly, the semi-major axis, the libration amplitude $A_{32}$, and the absolute H magnitude are all chosen following
the same procedure as for the non-Kozai plutinos.

\section{On-Sky Biases} \label{sec:biases}

We build up a population of synthetic plutinos, drawing orbital elements and magnitudes
from the specified distributions as described above,
and determine if each object is detected by a survey using the survey simulator code.  
Using the specified field coverage, magnitude efficiency,
and tracking fraction, the code determines
whether or not each object will be detected by the survey.
Comparing the distributions of the drawn and simulator-detected objects
gives an idea of the biases that are present in surveys that
cover different areas of the sky to different magnitude depths.
When the simulated detections are compared to the true detections in a real, 
well-characterized survey, this
is a powerful tool to help in debiasing to regain the real intrinsic population's
orbital distribution.

\begin{figure}
\centering
\includegraphics[scale=0.4]{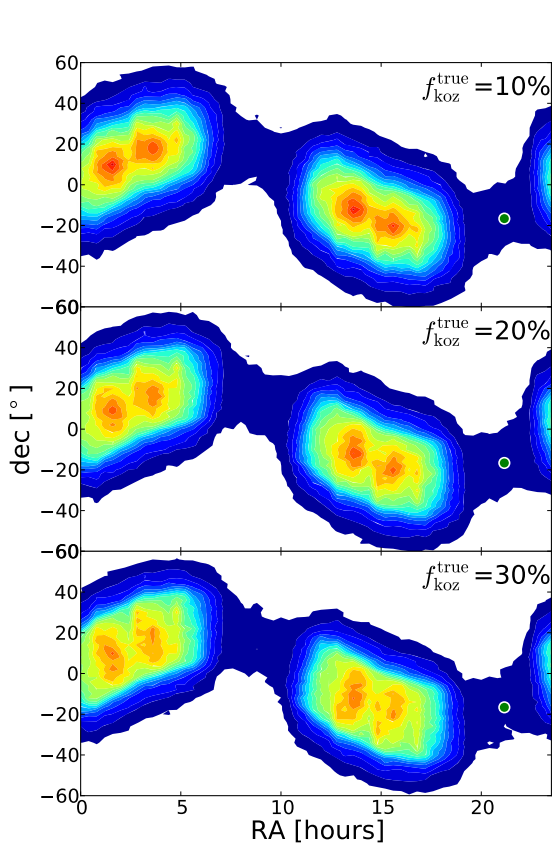}
\caption{
Relative density of detections on the sky in an all-sky survey using our plutino model, 
including both Kozai
and non-Kozai components, for three different values of $\fkoztrue$.
Contours are evenly spaced in detection density, and contour values are the same in all
three plots (absolute detection densities are arbitrary).
The position of Neptune is shown by a green circle.  
The detection density becomes somewhat less concentrated to the ecliptic for 
increasing Kozai fraction.
}
\label{fig:skydetections}
\end{figure}

Figure~\ref{fig:skydetections} shows an on-sky detection density map for all 
the plutinos, including the Kozai component.
With the Kozai component included, it is still true that
most plutinos are detected in broad clumps around the orthoneptune points,
90$\degree$ away from Neptune.
The reader will notice that the highest detection densities still occur in clumps
in the ecliptic
on either side of the orthoneptune points rather than exactly centered
on the orthoneptune points.
This is caused by the ``turnaround'' 
detection effect described in Section~\ref{sec:plutrealdist}.

The Kozai plutinos are only visible (for this realistic model) as subtle density enhancements about
10$\degree$ off the ecliptic, making the density contours in Figure~\ref{fig:skydetections}
appear slightly more rectangular than in Figure~\ref{fig:turnaroundfuzzy}.
This rectangular shape is enhanced for higher values of $\fkoztrue$.
Note that there is no ``spike'' in detections at the ecliptic latitudes where the 
density of Kozai plutinos is highest;
the detection densities are still dominated by the much greater numbers of non-Kozai plutinos.

\begin{figure}
\centering
\includegraphics[scale=0.4]{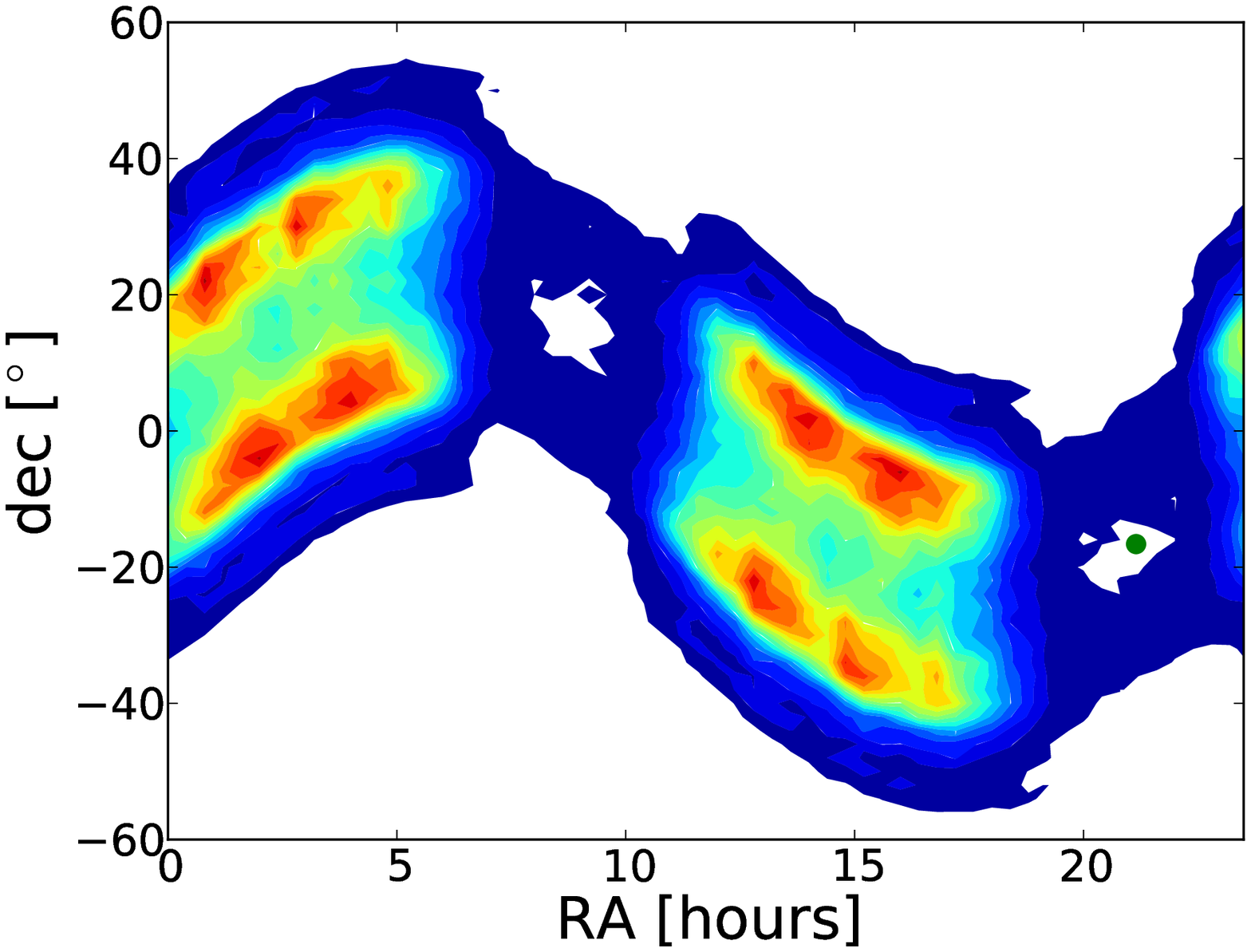}
\caption{Relative density of detections on the sky using an all-sky survey only using
the Kozai component of our model.
Contours are evenly spaced in detection density.
The position of Neptune is shown by a green circle.
}
\label{fig:kozdetections}
\end{figure}

To clarify what is happening for the Kozai population, 
the Kozai component is shown separately in Figure~\ref{fig:kozdetections}.
The Kozai plutino detections are more confined in ecliptic latitude than the non-Kozai plutinos,
with the highest detection densities happening about 10$\degree$ above and below the ecliptic,
in broad swaths surrounding the orthoneptune points, with the 
central minimum again caused by the lack of $A_{32}<20\degree$ plutinos.

\subsection{Ecliptic Latitude Distribution of Detections} \label{sec:eclipticlat}

\begin{figure}
\centering
\includegraphics[scale=0.4]{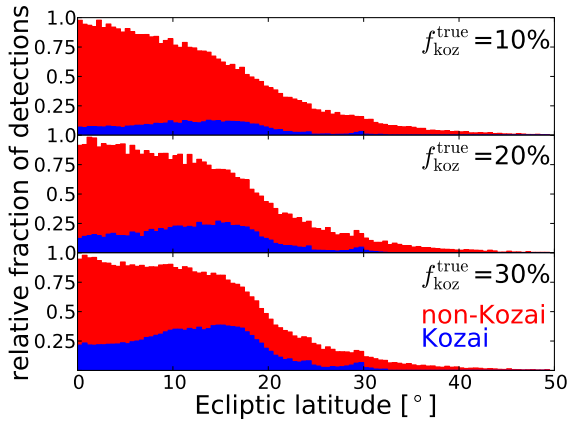}
\caption{
Stacked histograms of the ecliptic latitude distribution 
of detected Plutinos for different values of $\fkoztrue$,
using a magnitude-limited all-sky survey.
The number of detections per bin is normalized to the maximum bin.
Kozai plutinos are shown in red, non-Kozai plutinos in blue.
}
\label{fig:lat3plot}
\end{figure}

Figure~\ref{fig:lat3plot} presents the ecliptic latitudes of detected Plutinos in a simulated all-sky survey.
The number of detections for all plutinos smoothly falls from 0$\degree$ ecliptic latitude
on up to higher latitudes.
Although the number of Kozai detections climbs as one rises to $\sim$15$\degree$ ecliptic latitude,
they never hold more than about half the detections in a bin.

\citet{Schwambetal2010} and \citet{Brown2008} found a factor of $\sim$4 spike in 
the number of detections in their surveys in the 11-13$\degree$ ecliptic latitude bin, which they 
attribute to potentially being caused by Kozai plutinos.
However, our simulation makes this explanation implausible.
Even when we go to the extreme and unrealistic case of only using the lowest 
$\imax$ value of 14$\degree$
(which makes the Kozai plutinos as compact as possible in ecliptic latitude), and using
the highest value of $\fkoztrue=30\%$, we still find that there is only a $\sim$20\%
increase in the number of plutinos at ecliptic latitudes of 11-13$\degree$.
Kozai plutinos do not explain this spike in detections, because it is impossible
to confine the detections of Kozai librators to a narrow ecliptic latitude bin.
The reported detection spike is likely a small-number statistics fluctuation.

\subsection{$\fkoz$ On-Sky} \label{sec:fkozonsky}

\begin{figure}
\centering
\includegraphics[scale=0.32]{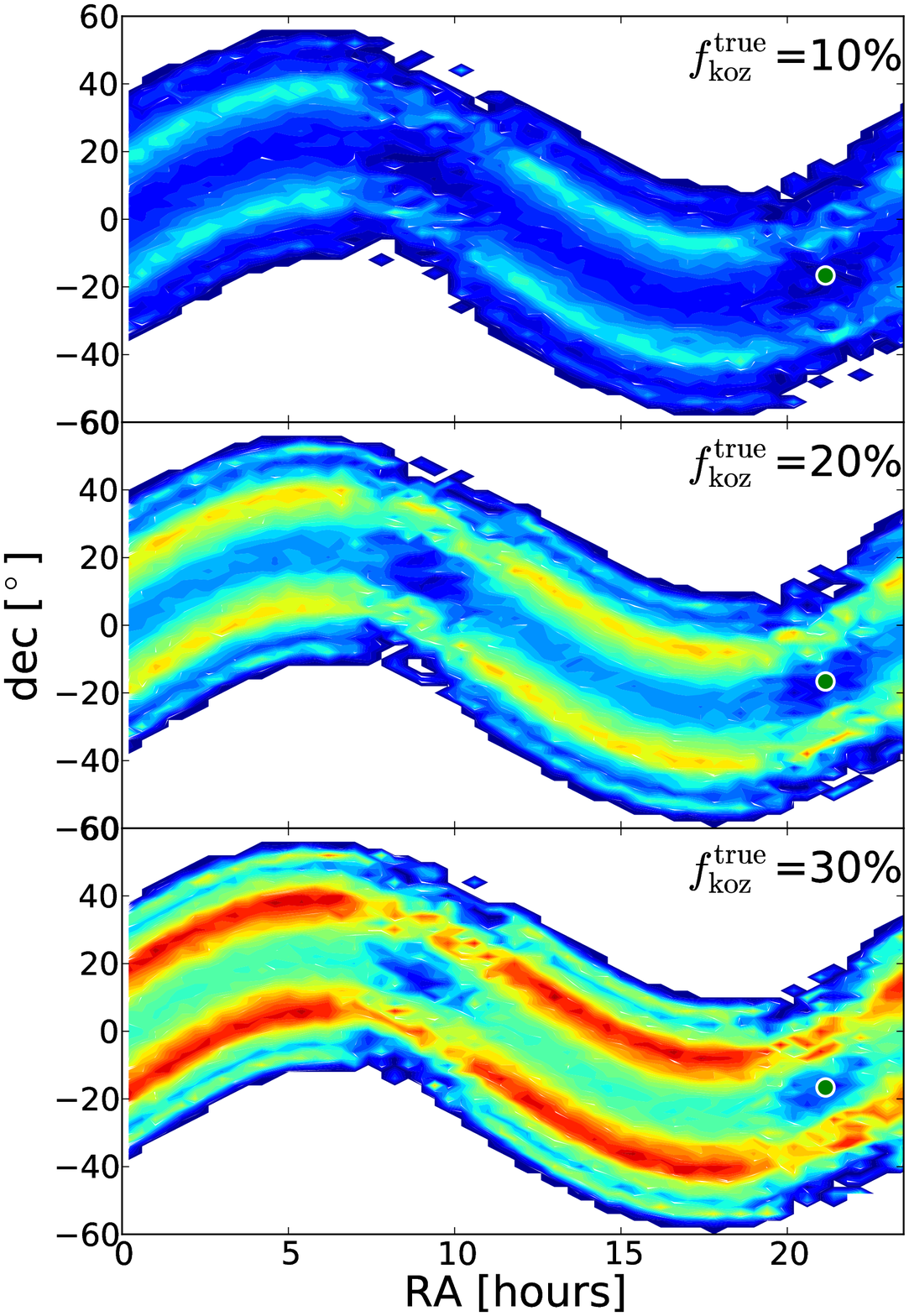} \includegraphics[scale=0.35]{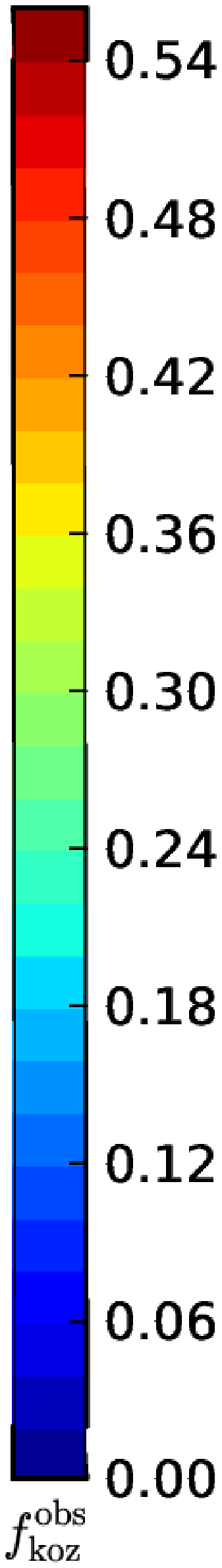}
\caption{This shows $\fkozobs$, the number of Kozai plutino detections divided
by the total number of plutino detections in $2\degree\times6\degree$ bins on the sky,
using our best plutino model.  
The true Kozai fraction (for the entire plutino population) $\fkoztrue$, 
is 10\% in the top panel, 20\% in the center, and 30\% in the lower panel.
The local $\fkozobs$ varies widely, from nearly zero around Neptune to 
its highest values about 10$\degree$ off the ecliptic.
}
\label{fig:fkoz}
\end{figure}

Figure~\ref{fig:fkoz} shows the ratio between the number of Kozai plutino detections to the
total number of plutino detections in small bins on the sky, providing a local $\fkozobs$ map.
(This is essentially the ratio of Figure~\ref{fig:kozdetections} to Figure~\ref{fig:skydetections}).
Figure~\ref{fig:fkoz} shows the range of $\fkozobs$ values that could be locally found at different positions
on the sky, for $\fkoztrue$ of 10\%, 20\%, and 30\%.

$\fkozobs$ values vary from 0\% to nearly twice the $\fkoztrue$ values.  
The highest $\fkozobs$ values occur where the Kozai detection density is highest:
above and below the ecliptic plane by about 12$\degree$

\subsection{Size Distribution Effects} \label{sec:alpha}

The diameter distribution of TNOs is fit by a power law, usually parameterized as
$N(>d) \propto d^{-Q}$, where $N(>d)$ is the number of objects larger than a 
diameter $d$, and $Q$ is the index of the power law.
However, because only a few of the largest KBOs have had their diameters directly
measured by occultation or resolved imaging, what is actually measured is the magnitude.
To convert this to a diameter, an albedo must be measured or assumed.
For this reason we discuss the size distribution in terms of absolute magnitude:
$N(<H)\propto10^{\alpha H}$.
The values $Q$ and $\alpha$ are related: $\alpha=\frac{Q}{5}$.

\begin{figure}
\centering
\includegraphics[scale=0.4]{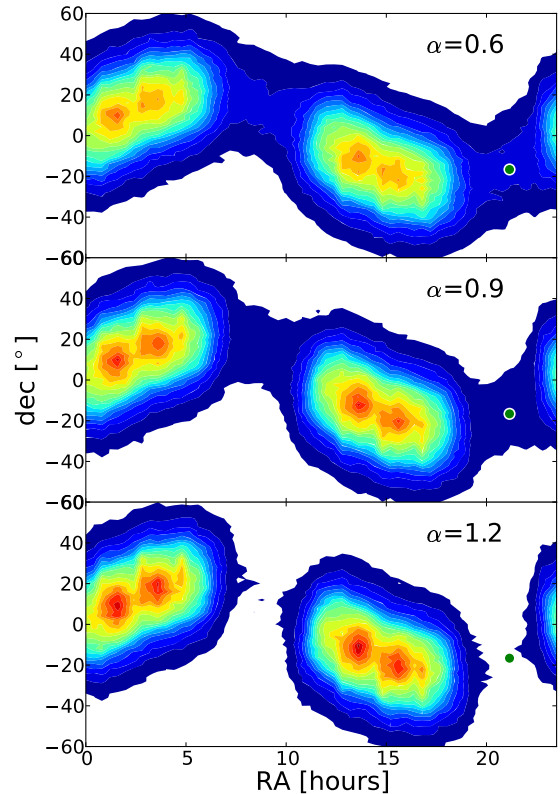}
\caption{Relative density of plutino detections (including Kozai, with $\fkoztrue=10\%$) using 
three different values of $\alpha$, the logarithmic slope of the
size distribution in absolute magnitude.
Contours are evenly spaced in detection density, and the contours represent the same
values in each panel.
For steeper power laws, the density of detection also becomes steeper, with
more detections at the turnaround points, and fewer detections 90$\degree$ away.
}
\label{fig:plutdetections_alphas3}
\end{figure}

\begin{figure}
\centering
\includegraphics[scale=0.32]{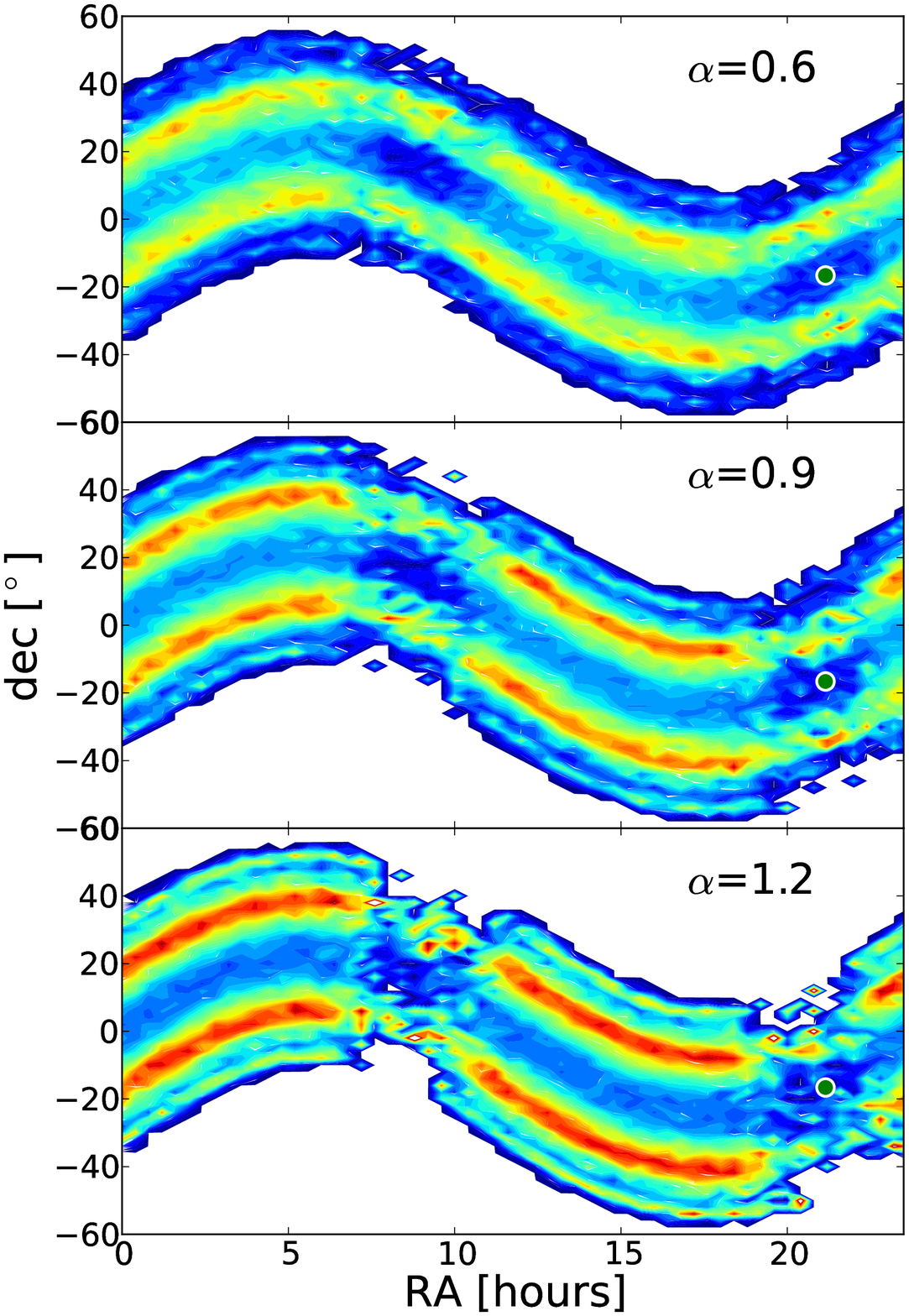} \includegraphics[scale=0.35]{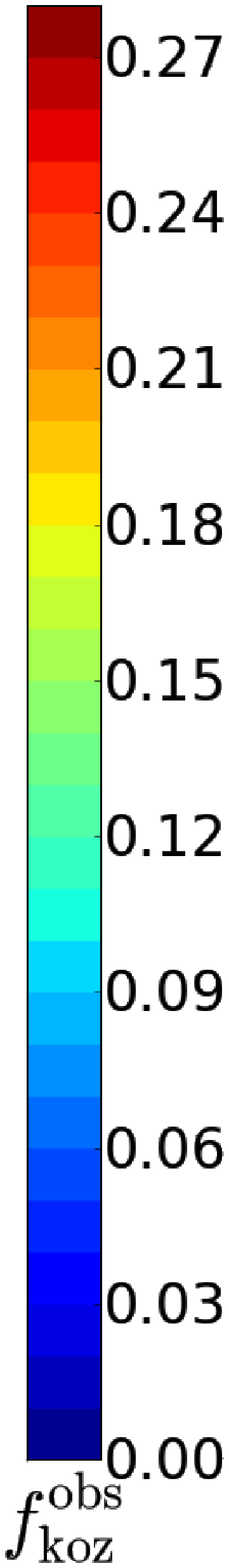}
\caption{$\fkozobs$ at different points on the sky, using $\fkoztrue=10\%$, and
three different slopes for the power-law absolute magnitude distribution.
Steeper slopes cause the $\fkozobs$ value to vary more widely, reaching maximum values of almost 30\%,
while the shallow slopes give the highest $\fkozobs$ values of only about 20\%.
}
\label{fig:fkoz_alphas3}
\end{figure}

The logarithmic slope $\alpha$ is known to be different depending on the size of the KBOs
\citep[e.g.][]{FuentesHolman2008, FraserKavelaars2009}.
For most of our calculations, we use the nominal CFEPS value for plutinos of $\alpha=0.9$
\citep{GladmanReson}.
But for comparison, Figure~\ref{fig:plutdetections_alphas3} shows the effect
of different values of $\alpha$ on the detection density,
and Figure~\ref{fig:fkoz_alphas3} shows the effect of different $\alpha$ values
on $\fkozobs$ at different places on the sky.
Steeper slopes result in steeper
detection density distributions, where the peak detection densities are
much higher.
This is because the relative importance of detecting the large number of small
plutinos that are only visible at perihelion increases.
Lower values of $\alpha$ result in shallower density distributions.
This effect is noticeable, but the overall pattern of where on the sky the highest $\fkozobs$
values are remains the same.

\subsection{Example Simulated Surveys} \label{sec:simsurveys}

We perform a number of strawman simulated surveys to demonstrate the different values of 
apparent average $\fkozobs$ 
that result from different survey parameters and different $\fkoztrue$ values.
These are summarized in Table~\ref{tab:fkozfake}. 
For all of these simulated surveys, we ignore the extra confusion caused by the plane of the Milky Way, 
and assume that 
the entire area within each survey is observed uniformly and with perfect tracking efficiency
(that is, all discoveries are tracked to yield high quality orbits).

\begin{deluxetable*}{rl|ccc} 												
\tabletypesize{\small}												
\tablecaption{Simulations of $f_{koz}$												
\label{tab:fkozfake}}												
\tablehead{												
	&	Survey Description	&	\multicolumn{3}{c}{$\fkozobs$ values for:}								\\
	&		&	$\fkoztrue=10\%$		&	$\fkoztrue=20\%$		&	$\fkoztrue=30\%$		}
\startdata												
1	&	all sky 	&	12	\%	&	23	\%	&	35	\%	\\
2	&	5$\degree$ high, on ecliptic	&	7.5	\%	&	15	\%	&	23	\%	\\
3	&	20$\degree\times$20$\degree$ box, 90$\degree$ from Neptune, on ecliptic	&	9.5	\%	&	19	\%	&	29	\%	\\
4	&	20$\degree\times$20$\degree$ box, 75$\degree$ from Neptune, on ecliptic	&	10	\%	&	18	\%	&	28	\%	\\
5	&	2$\degree\times$2$\degree$ box, 90$\degree$ from Neptune, on ecliptic	&	7.6	\%	&	16	\%	&	24	\%	\\
6	&	2$\degree\times$2$\degree$ box, 75$\degree$ from Neptune, on ecliptic	&	7.4	\%	&	16	\%	&	23	\%	\\
7	&	20$\degree\times$20$\degree$ box, 90$\degree$ from Neptune, 10$\degree$ off ecliptic	&	13	\%	&	25	\%	&	37	\%	\\
8	&	20$\degree\times$20$\degree$ box, 75$\degree$ from Neptune, 10$\degree$ off ecliptic	&	13	\%	&	24	\%	&	36	\%	\\
9	&	2$\degree\times$2$\degree$ box, 90$\degree$ from Neptune, 10$\degree$ off ecliptic	&	14	\%	&	26	\%	&	37	\%	\\
10	&	2$\degree\times$2$\degree$ box, 75$\degree$ from Neptune, 10$\degree$ off ecliptic	&	14	\%	&	27	\%	&	38	\%	\\
\enddata												
\end{deluxetable*}												

Each simulated survey goes to a magnitude depth of $g=24.9$.
Varying the magnitude depth did not have any noticeable effect on
the detection density or $\fkoz$ values on the sky, due to the 
assumed exponential nature of the distribution.

First we perform an all-sky survey with a set limiting magnitude (Survey~1),
which finds a higher Kozai fraction than reality, with
$\fkozobs$ being higher than $\fkoztrue$.
Survey~2, an ecliptic survey covering the entire ecliptic within $\pm2.5\degree$,
unsurprisingly finds the opposite effect, with $\fkozobs$ being lower than $\fkoztrue$.
This is due to the Kozai plutinos preferentially being detected out of the ecliptic plane,
as discussed in Section~\ref{sec:kozonsky}.

Surveys~3 and 4 are 20$\degree\times$20$\degree$ surveys centered on the ecliptic.  
Survey~3 is centered 90$\degree$ away from Neptune, and Survey~4
is centered 75$\degree$ away from Neptune, near the plutinos' peak in detectability.
Surveys~5 and 6 are smaller, 2$\degree\times$2$\degree$ surveys, centered on 
the ecliptic 90$\degree$ and 75$\degree$ away from Neptune, respectively.
These four surveys all find lower $\fkozobs$ than $\fkoztrue$, for the same
reason as Survey~2.

Surveys~7 and 8 are the same as Surveys~3 and 4, except centered 10$\degree$ above
the ecliptic.  
Similarly, Surveys~9 and 10 are the same as Surveys~5 and 6, raised to 10$\degree$
above the ecliptic.
Because these surveys cover the range of the Kozai plutinos' peak detection
density, they all find higher $\fkozobs$ than $\fkoztrue$ values.

There is not a significant difference between the $\fkozobs$ values measured by the 
surveys that are centered on 90$\degree$ from Neptune and those
centered on 75$\degree$ from Neptune.
There {\it is} a difference in the relative number of detections, with overall more plutinos
detected at 75$\degree$ from Neptune.  
However, because both the Kozai and non-Kozai plutinos have the same $A_{32}$ 
distribution in this model, the Kozai fraction does not vary significantly between these
two positions on the sky.

These surveys demonstrate that very different values of $\fkozobs$ can be measured
depending on the on-sky location of the survey. 
Due to the different biases inherent in the distribution of Kozai plutinos 
versus non-Kozai plutinos,
careful debiasing is required to calculate $\fkoztrue$ from any survey,
even one which covers the entire sky.

\section{Comparison with Previous Literature} \label{sec:fkoz}

In the previous sections, we have discussed two quantities that can be measured
for a survey or simulation that contains many well-characterized plutinos: 
$\fkoz$ and the distribution of $\imax$ for the Kozai plutinos.
Below, we discuss these quantities as measured by observational surveys
and giant planet migration simulations.
Only a few surveys are discussed here,
as only a few previously published TNO surveys have rigorous enough tracking and characterization methods to 
classify plutinos as Kozai or non-Kozai in orbital integrations.

\subsection{The Kozai Fraction $\fkoz$}

The simplest quantity to measure in a survey or simulation that contains plutinos
is $\fkoz$, the fraction of plutinos that are in Kozai.
However, one must be careful to note whether this is the true or apparent $\fkoz$.
Most surveys will have some bias, as shown in Table~\ref{tab:fkozfake}, 
causing $\fkozobs$ to be different than $\fkoztrue$.

Below we discuss the $\fkoz$ results presented in several observational surveys
and theoretical simulations.
A summary is presented in Table~\ref{tab:fkoz}.

\begin{deluxetable*}{l|lccc}									
\tabletypesize{\small}									
\tablecaption{Measurements of $\fkoz$ from the Literature									
\label{tab:fkoz}}									
\tablehead{									
Source	&	type	&	$\fkoztrue$	&	$\fkozobs$	&	\# plutinos	}
\startdata									
\citet{Gomes2000}	&	Observational	&	-	&	26\%	&	23	\\
\citet{Nesvornyetal2000}	&	Observational	&	-	&	12\%	&	33	\\
\citet{ChiangJordan2002}	&	Theoretical	&	20-30\%	&	-	&	42	\\
\citet{HahnMalhotra2005}	&	Theoretical	&	19\%	&	-	&	133	\\
\citet{LykawkaMukai2007}	&	Observational	&	-	&	22-30\%	&	100	\\
\citet{Levisonetal2008}	&	Theoretical	&	16\%	&	-	&	186	\\
\citet{Schwambetal2010}	&	Observational	&	-	&	33\%	&	6	\\
\citet{GladmanReson}	&	Observational	&	10\%	&	8\%	&	24	\\
\enddata									
\end{deluxetable*}									

\subsubsection{Observational Surveys: $\fkozobs$}

CFEPS \citep{PetitL7}, being a well-calibrated survey, was able to provide both
an apparent and a true $\fkoz$, albeit with large uncertainty \citep{GladmanReson}.
They find $\fkozobs$ of $2/24=8\%$.
After debiasing, this would require a value of $\fkoztrue$ of 10\%.
However, because CFEPS was confined to the ecliptic plane, it was not very sensitive to the 
high-inclination Kozai plutino population, and 
$\fkoztrue$ up to 33\% cannot be ruled out with 99\% confidence
due to the small number statistics of having only two detected Kozai plutinos.

The Deep Ecliptic Survey \citep{Elliotetal2005}, while finding a reported 51 plutinos,
did not specifically label any of their discovered plutinos as Kozai, 
and thus is not discussed further (although some of their discoveries are in the 
biased \citet{LykawkaMukai2007} compilation, discussed below).

A few papers have tried to use the entire MPC database as a survey.
While this does provide many plutinos, the MPC contains the results of many surveys and
even serendipitous discoveries, each with completely different and possibly unknown
biases, since one doesn't know where searches failed to detect plutinos.  
Debiasing $\fkozobs$ to find $\fkoztrue$ is impossible for these surveys.

\citet{Gomes2000} and \citet{Nesvornyetal2000} performed similar large MPC database searches capable of 
classifying objects as Kozai or non-Kozai plutinos.  
\citet{Gomes2000} examined the first 23 discovered Plutinos with observations for 2 or more oppositions.  
Though many of these classifications were provisional due to a lack of precise data, 
he found $\fkozobs$ of $6/23=26\%$.
\citet{Nesvornyetal2000} performed a similar analysis for the first 33 plutinos,
finding that only 4 of them were in Kozai, giving $\fkozobs$ of 12\%,
despite an overlapping sample.

Currently, the largest collection of well-classified plutinos was presented in \citet{LykawkaMukai2007},
with 100 plutinos from the MPC database.
All of these plutinos had at least 2 opposition observations, and 10~Myr orbital integrations were performed.
They found that 22 plutinos are solidly in the Kozai resonance, with 8 more that are in the Kozai resonance for 
part of their integration.
Thus, from their integrations they find $\fkozobs$ of 22-30\%.

\citet{Schwambetal2010} completed a wide-field survey covering a large fraction of the sky 
($\sim12,000$ square degrees)
within $30\degree$ of the ecliptic.
This relatively shallow survey ($R\sim21.5$) found 6 plutinos, two of which are in Kozai, giving 
$\fkozobs=33\%$, albeit with large Poisson uncertainty.
The higher ecliptic latitudes included in this survey would make detecting the Kozai plutinos 
more likely, thus this higher apparent $\fkozobs$ value is not surprising.
Although this is the first large area survey which found and tracked plutinos and Kozai plutinos, 
a much larger number of plutino detections will be needed to accurately measure the Kozai fraction.

\subsubsection{Theoretical Simulations: $\fkoztrue$}

Of the published simulations of giant planet migration, only \citet{ChiangJordan2002}
includes information on which plutino test particles ended up in Kozai.  
Future simulations should include this information, as it may prove a useful diagnostic.
We also discuss the Kozai plutinos from 
\citet{HahnMalhotra2005} and \citet{Levisonetal2008}, because the authors
provided us with the output orbits of these simulations and were able to complete 
the required analysis ourselves.

\citet{ChiangJordan2002} studied a smooth outward migration of Neptune, with different migration times
for Neptune to reach its current location.
They discuss objects that are captured into the Kozai resonance within the 3:2 for their
simulation where Neptune migrates with a damping half-life of 10$^7$ years.
Because of the shorter timescale of their simulations, their resonance classification isn't 
as secure as in the other simulations discussed below.  
Out of 92 plutinos at the end of their simulation, they estimate that 42 will remain in the
3:2 resonance for the age of the solar system.  
Of these, 8-12 are in Kozai, giving $\fkoztrue$ of 20-30\%.
They unfortunately do not discuss the effect that different migration timescales have
on the Kozai fraction, nor how $\fkoztrue$ might evolve over 4~Gyr.

\citet{HahnMalhotra2005} and \citet{Levisonetal2008} provided enough data from the end of their
theoretical migration simulations that we were able to continue the integrations for 10~Myr, long enough to 
diagnose if a plutino is in Kozai or not.
\citet{HahnMalhotra2005} used a smooth outward migration of Neptune, while \citet{Levisonetal2008}
had Neptune on a large-eccentricity orbit that damps after interacting with the Kuiper Belt
\citep[motivated by the ``Nice Model'' scenario;][]{Tsiganisetal2005}.

For the \citet{Levisonetal2008} simulation, we were given the 10~Myr orbital integrations originally
used to classify objects as resonant or non-resonant at the end of their 1~Gyr migration simulation (Run~B).
These integrations contain the osculating orbital elements at each timestep, and these were searched 
for oscillation of $\omega$ around 90$\degree$ or 270$\degree$ to determine $A_{\omega}$.

For the \citet{HahnMalhotra2005} data, we were provided the osculating orbital elements of all test particles
and the 4 giant planets at the end of their 4.5~Gyr giant planet migration simulation. 
However, these were divided into 100 separate simulations, each with different giant planet positions and different
numbers of remaining test particles (most had $\sim$50).  
Each of these were input into a slightly modified version of 
the orbital integrator SWIFT \citep{LevisonDuncan1994}, 
and 30~Myr orbital integrations were performed.  
We analysed the remaining test particles for libration of $\phi_{32}$, and then for oscillation of 
$\omega$ around 90$\degree$ or 270$\degree$.

Out of 133 plutinos in the \citet{HahnMalhotra2005} simulation, 25 were in Kozai, giving 
$\fkoztrue=19\%$.
The \citet{Levisonetal2008} simulation provided 186 plutinos, of which 29 were in Kozai,
giving $\fkoztrue=16\%$.

These $\fkoz$ values all contain large uncertainties, and in our opinion, all 
roughly agree with each other at this point.  
As more plutinos are found by rapidly repeating all-sky surveys such as LSST,
the value of $\fkoztrue$ should become precisely measurable as the survey characterization
becomes well-determined.

\subsection{Distribution of Kozai Parameters} 

\begin{figure}
\centering
\includegraphics[scale=0.4]{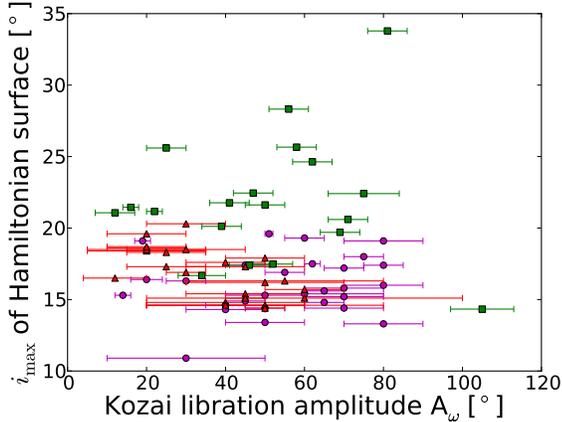}
\caption{Scatterplot showing the $\imax$ values (which parameterize which Hamiltonian level surface the 
plutino is on), and the Kozai libration amplitudes $A_{\omega}$ (which parameterizes which contour of the level surface
the plutino is on).
Purple circles show simulated Kozai plutinos from \citet{Levisonetal2008} (LMVGT2008), 
green squares show real Kozai plutinos from the MPC database \citep[L\&M2007,][]{LykawkaMukai2007},
and red triangles show simulated Kozai plutinos from \citet{HahnMalhotra2005} (H\&M2005).
Error bars for the MPC values are those reported by \citet{LykawkaMukai2007}, other error
bars are estimated by eye from orbital integrations.
}
\label{fig:models_mpc}
\end{figure}

\begin{figure}
\centering
\includegraphics[scale=0.4]{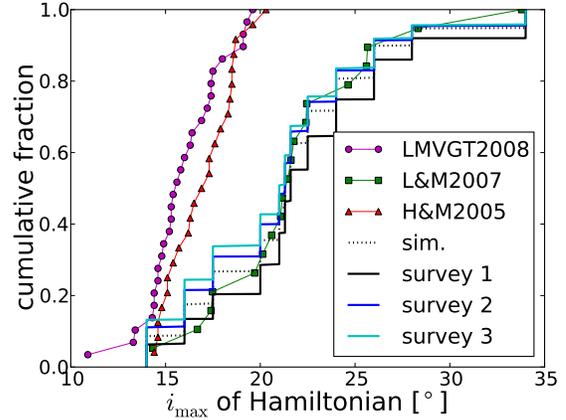} 
\includegraphics[scale=0.4]{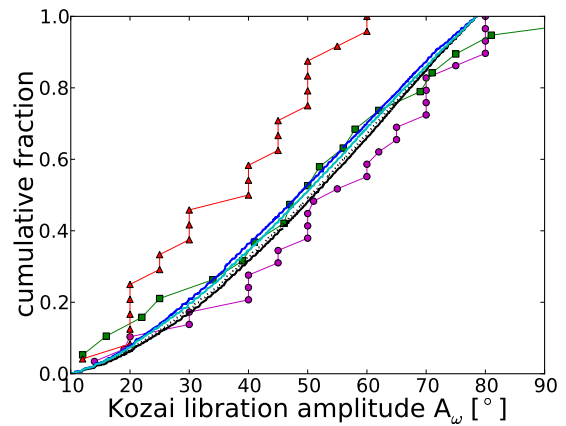}
\caption{
Cumulative distributions of $\imax$ and $A_{\omega}$ vales.
See Figure~\ref{fig:models_mpc} caption for symbols.
The intrinsic distribution of thousands of simulated Kozai plutinos using our model is shown ‌
by a dotted black line.
The results of running this distribution through 
a survey simulator using parameters for Surveys~1, an all-sky survey, Survey~2, an ecliptic survey, and 
Survey~3, a 20$\degree\times$20$\degree$ survey centered 90$\degree$ from Neptune (see Section~\ref{sec:simsurveys})
are shown by solid lines.
The models produce generically lower inclination objects, but seem to produce similar Kozai
libration amplitudes to the MPC objects (keep in mind that the MPC objects are not debiased).
Note that the intrinsic model matches the MPC sample quite well because, 
as explained in Section~\ref{sec:pop:kozai},
the distribution of Kozai $\imax$ values was based on the objects in the MPC.
}
\label{fig:models_mpc2}
\end{figure}

The two main parameters we use to describe the Kozai behavior of a given Kozai plutino
are $\imax$ and the Kozai libration amplitude $A_{\omega}$.
The distribution of $\imax$ tells about the range of $e$ and $i$ that are possible during a
Kozai cycle.
$\imax$ is a parameterization of which Hamiltonian level surface currently best describes the Kozai
libration of that plutino.  
The Kozai libration amplitude $A_{\omega}$ measures which contour within the $\imax$ level 
surface the plutino
is on, and is found from looking at the results of a 10-30~Myr diagnostic orbital integration.

Because this is a distribution and not just a single value like $\fkoz$, it is only instructive
to analyse for the surveys and simulations with the largest number of plutinos.
We compare the $\imax$ distributions found in the theoretical giant planet migration 
simulations of \citet{HahnMalhotra2005}
and \citet{Levisonetal2008} with the MPC database analysis presented in \citet{LykawkaMukai2007},
and with our own simulation (Figures~\ref{fig:models_mpc} and \ref{fig:models_mpc2}).

When looking at Figures~\ref{fig:models_mpc} and \ref{fig:models_mpc2}, it is imporant to keep in mind that we are comparing 
different kinds of distributions.
The simulated Kozai plutinos from \citet{Levisonetal2008}, \citet{HahnMalhotra2005}, and this paper 
are intrinsic distributions, that is, not observed by a biased survey.
The MPC-detected Kozai plutinos \citep[from][]{LykawkaMukai2007} 
and the distributions resulting from the simulated surveys presented 
in this paper {\it are} biased.
In the case of the MPC sample, which contains the results of many uncharacterized
surveys, precise debiasing is impossible.
The simulated surveys, however, are all based on our simulated plutino distribution, so we can
see the effects of different types of surveys on the detected parameters.
The all-sky survey (Survey~1) is slightly biased toward finding a higher proportion of higher $\imax$
objects than reality, while Surveys~2 and 3 are weakly biased toward finding a higher 
proportion of lower $\imax$
objects than reality.
All three simulated surveys show little bias in the distribution of Kozai libration amplitudes.

The distribution of Kozai libration amplitudes is shown in the bottom panel of Figure~\ref{fig:models_mpc2}.
There is general agreement between libration amplitude distribution of the MPC Kozai plutinos
and both giant planet migration simulations,
however, the \citet{HahnMalhotra2005} simulation finds generally lower libration amplitudes,
while the \citet{Levisonetal2008} simulation finds generally higher.
This is an area that could use more theoretical work, as different migration timescales
and migration modes
may cause different libration amplitude distributions.

It is obvious from the top panel of Figure~\ref{fig:models_mpc2}
that the $\imax$ values are much lower in both of the simulations than in the MPC.
Although the MPC distribution is not debiased, and thus may not reflect the true distribution of 
Kozai plutinos, the inclination distribution discrepancy between models and the true Kuiper Belt
has been noticed before \citep{GladmanReson}.
This is a generic problem with giant planet migration simulations, and not unique to this Kozai problem:
these simulations are not good at raising the inclinations of the captured resonant objects 
\citep[noted by][and others]{ChiangJordan2002}.

\section{Conclusion} \label{sec:conclusion}

With the upcoming inauguration of such rapid-fire all-sky surveys as 
LSST \citep{LSSTbook,Jonesetal2009} and Pan-STARRS \citep{Gravetal2011}, 
which are expected to detect hundreds of new TNOs, 
we are entering an era when we have enough well-characterized
plutinos to be able to debias and measure the value of $\fkoztrue$ with
more precision than has been possible.

Little theoretical work has been done relating the value of $\fkoztrue$ to
the migration timescale of Neptune, but this may be an important and helpful diagnostic.
To our knowledge, no theoretical work has been done so far to understand how 
the $\imax$ distribution is set, and how it evolves over time.
There are also other relationships that have not been explored, such as 
the relation between the libration amplitude of $\phi_{32}$ and the Kozai libration
amplitude.

Our results allows optimization for observers planning targeted surveys.
If the goal of the survey is to find as many plutinos as possible,
the highest density on the sky is not exactly 90$\degree$ away from Neptune, but about
15$\degree$ on either side of $\lambda_N\pm90\degree$.
If the goal of the survey is to find as many Kozai plutinos as possible,
the best places on the sky are about 12$\degree$ above and below the ecliptic,
and 15$\degree$ on either side of the orthoneptune points.
The value of $\fkozobs$ that is measured in a given survey can be significantly
different from $\fkoztrue$, and careful debiasing is necessary to derive
the true value.
Parameters of the survey such as pointings, field depths, tracking efficiencies,
and fields with no detections must all be characterized in order to properly
debias the results \citep[see][]{Jonesetal2010}.

\acknowledgements{
The authors wish to thank C.~Van~Laerhoven, H.~Levison, and J.~Hahn for providing
us with output from their migration simulations, and X.-S.~Wan and T.-Y.~Huang for 
providing us with the disturbing function coefficients for Kozai plutinos.
}


\end{document}